\documentclass[aps,pre,reprint,floatfix]{revtex4-1}
\usepackage{blindtext}
\usepackage{lipsum}  
\usepackage{amsmath}
\usepackage{amsfonts}
\usepackage{amssymb}
\usepackage{comment}
\usepackage{graphicx}
\usepackage{dcolumn}
\usepackage{bm}
\usepackage{physics}
\usepackage{xcolor}
\usepackage{caption}
\usepackage{subcaption}
\usepackage{hyperref}
\usepackage{tikz}

\newcommand{\spt}[1]{\textcolor{black}{#1}}

\begin{document}
\title{Simulating dynamics of ellipsoidal particles using lattice Boltzmann method}
\author{Sumesh P Thampi$^1$, Kevin Stratford$^2$, Oliver Henrich$^3$}
 \affiliation{$^1$ Department of Chemical Engineering, Indian Institute of Technology Madras, Chennai-36, India\\ $^2$  EPCC, The University of Edinburgh, Edinburgh EH8 9BT, United Kingdom\\$^3$ Department of Physics, University of Strathclyde, Glasgow G4 0NG, United Kingdom}

\begin{abstract}
Anisotropic particles are often encountered in different fields of soft matter and complex fluids. In this work, we present an implementation of the coupled hydrodynamics of solid ellipsoidal particles and the surrounding fluid using the lattice Boltzmann method. A standard link-based mechanism is used to implement the solid-fluid boundary conditions. We develop an implicit method to update the position and orientation of the ellipsoid. This exploits the relations between the quaternion which describes the ellipsoid's orientation and the ellipsoid's angular velocity to obtain a stable and robust dynamic update. The proposed algorithm is validated by looking at four scenarios: (i) the steady translational velocity of a spheroid subject to an external force in different orientations, (ii) the drift of an inclined spheroid subject to an imposed force, (iii) three-dimensional rotational motions in a simple shear flow (Jeffrey's orbits), and (iv) developed fluid flows and self-propulsion exhibited by a spheroidal microswimmer. In all cases the comparison of numerical results \spt{shows} good agreement with known analytical solutions, irrespective of the choice of the fluid properties, geometrical parameters, and lattice Boltzmann model, thus demonstrating the robustness of the proposed algorithm.
\end{abstract}

\maketitle
\section{Introduction}
The approximation that the shape of an object is a sphere is often a first step made to solve problems in physics (occasionally to the point of caricature \cite{stellman1973}). Such an approximation allows the development of simple analytical solutions in a variety of problems. Examples include the calculation of the Stokes drag for viscous flow around a spherical particle in fluid mechanics \cite{batchelor1967introduction}, and determining the scattering cross section due to a spherical particle impinged upon by an electromagnetic wave \cite{jackson2021classical}. The effect of shape anisotropy may then understood by performing perturbation theory to the spherical particle approximation, or considering the opposite limit of an infinitely long particle (slender body limit). Calculations may require rather non-conventional and cumbersome coordinate systems. Interestingly, numerical simulations also use a spherical particle approximation in some circumstances since it helps (i) validate the simulation method accurately and (ii) exploit the symmetry in the system to reduce the computational load. However, it is also known that shape of objects, in and of itself, may lead to interesting physics as illustrated in the examples below. In this work, we discuss the implementation of the lattice Boltzmann method to simulate the dynamics of \spt{rigid} spheroidal particles immersed in a fluid and demonstrate the reliability of the method by comparing it with known analytical solutions. \spt{The study focuses on fluid mechanical consequences of shape anisotropic particles in low Reynolds number with intended applications in soft and biological matter.}

Fluid dynamics associated with moving objects has always been an area of interest to both scientists and engineers alike. Analytical solutions, even in the simplest geometries such as flow past a spherical particle, are often not available owing to the presence of nonlinear terms in the governing Navier-Stokes equations. Hence, the field of computational fluid dynamics has grown over several decades with different formulations based on finite difference, finite volume or finite element methods, developed primarily for engineering applications. At the same time, particle based methods such as dissipative particle dynamics, lattice Boltzmann simulations and multi-particle collision dynamics have been developed to investigate the science of mesoscale systems often encountered in the field of soft and biological matter \cite{desplat2001ludwig, harting2005large, aidun2010lattice, kruger2011efficient, kruger2017lattice}. The appeal of particle based methods is the relative simplicity of the numerical method, the more natural correspondence with the atomic or mesoscale picture rather than a continuum and hence the ability to capture features like thermal fluctuations. 

In particular, the lattice Boltzmann method (LBM) has been shown to be successful in simulating colloidal dispersions, for example dynamics of single and multiple rigid colloids in fluid flows \cite{ladd1994numerical, stratford2005colloidal, aidun2010lattice, thota2023numerical}, dynamics of microswimmer suspensions \cite{lintuvuori2017hydrodynamics, stenhammar2017role}. Most of these investigations have dealt with spherical particles. However, synthesis of non-spherical colloidal particles have become more common, and their dispersions exhibit emergent properties \cite{glotzer2007anisotropy, sacanna2011shape, anjali2018shape}. Microswimmers exploit their anisotropic shape to overcome the limitations imposed at low Reynolds number to self-propel \cite{lauga2020fluid}. Hence, shape anisotropy of colloidal and active particles is rather a common theme in the evolving scientific literature, but relatively less work has been carried out to simulate the dynamics of non-spherical objects dispersed in a fluid using the lattice Boltzmann method.

Moreover, the lattice Boltzmann method has been found to be successful in simulating a variety of mesoscale systems: droplet dynamics and interfacial flows \cite{pooley2008contact, kusumaatmaja2007modeling,  christianto2022modeling, harting2005large}, wetting and coating flows \cite{ledesma2011controlled}, surfactant adsorption kinetics \cite{zong2020modeling, van2018lattice, farhat2011hybrid}, nematic and cholesteric liquid crystal dynamics \cite{henrich2011structure, marenduzzo2007steady, cates2009lattice} and active fluids \cite{stenhammar2017role, marenduzzo2007steady, cates2009lattice, saw2017topological, rorai2022coexistence, fadda2017lattice, carenza2019lattice}. The common theme among these systems is the presence of additional governing equations for the order parameters that describe the microstructure of the material under consideration. These investigations have illustrated that the lattice Boltzmann method correctly captures the microstructure evolution coupled with hydrodynamics. The field of hyper-complex fluids that are mixtures of more than one type of constituent is an emerging area \cite{dogic2014hypercomplex, zhou2014living, lintuvuori2017hydrodynamics, stratford2005colloidal, lesniewska2022controllable, lintuvuori2018mixtures, blow2014biphasic, gidituri2021dynamics}. Systems with shape anisotropic particles at fluid-fluid interfaces, in liquid crystals as well as active and biological fluids are being investigated. Simulating such hyper-complex fluids as a prospective application, a robust lattice Boltzmann method that can handle non-spherical particles is desirable.

Dynamics of ellipsoidal particles in the framework of the lattice Boltzmann method has been studied in the literature. \citet{huang2012shear} analyzed the response of a spheroidal particle subjected to a Couette flow and hence determined the shear viscosity of dilute suspension of spheroidal particles. Since the spheroidal particle was placed at the center line in the Couette flow setup where the fluid velocity is zero, accounting for the translation of the center of mass of the spheroids was missing in this work. Similarly, the \spt{flow induced rotational} dynamics of particles is absent in studies such as \cite{rong2015lattice, ezzatabadipour2017fluid, dietzel2016application, trunk2021study,eshghinejadfard2016direct,holzer2009lattice} where the flow past a cluster of spherical \spt{or nonspherical} particles is investigated. In studies where the dynamics of particles is considered, it is typically restricted to spherical particles.  However, it is important to choose a stable numerical scheme that couples the translational and rotational motion of the non-spherical particles combined with lattice Boltzmann algorithm. This is particularly important when accounting for hydrodynamic interactions in non-dilute suspensions, hydrodynamic interactions with confining walls, as well as with regard to the effects arising from the `complex' nature of the fluids. 
Hybrid methods such as a combination of lattice Boltzmann and immersed boundary method (IBM) are also proving to be reliable tools. Studies using coupled LBM-IBM or similar approaches \cite{karimnejad2018sedimentation, wang2022numerical, hui2023sedimentation, kohestani2019non} for the sedimentation of non-spherical particles are analyzed, but restricted to two-dimensions where the rotation of the particles is less difficult to model. Even in more generalized investigations with three-dimensional simulations, the rotation of the spheroidal particles is restricted to two dimensions \cite{livi2021influence, zhao2023metaball}. The dynamics of prolate and oblate spheroidal particles in three dimensions is analyzed by \cite{eshghinejadfard2017fully, eshghinejadfard2018lattice}. However, the focus of these studies was on turbulent flows with large particle volume fractions. Thus, the effectiveness of the method in capturing single particle dynamics, whether passive or active, is less clear. A more detailed study is given by \cite{huang2014sedimentation} where the dynamics of ellipsoidal particles in narrow tubes is analyzed. Methods are also being developed that mesh the particles separately and differently from the conventional LBM Cartesian grid \cite{romanus2021immersed, romanus2022fully}.

\spt{To summarize, a variety of methods and numerical techniques are available in the literature to simulate the dynamics nonspherical particles, but a stable lattice Boltzmann algorithm that simulates the coupled translational, rotational dynamics of rigid ellipsoidal particles in three dimensions along with associated fluid mechanics is required to be developed. Currently, the available studies in the literature focus only on limited aspects of this problem as described above. This is particularly important considering the potential that lattice Boltzmann method offers to simulate dynamics of hyper-complex fluids containing rigid, nonspherical particles. Hybrid methods such as IBM-LBM are more suitable for elastic particles \cite{kruger2014deformability,delouei2022direct}; coupling the order parameter equations of complex fluids with hybrid methods is tedious and not demonstrated so far.}
Therefore, devising a robust, stable numerical algorithm to simulate the dynamics of ellipsoidal particles with lattice Boltzmann method is important, which is the theme of this paper. 
To this end we restrict ourselves to the analysis of the dynamics of active and passive prolate spheroidal particles dispersed in a Newtonian fluid, particularly focusing on capturing various well-known low Reynolds number phenomena, \spt{with intended applications in soft matter and complex fluids}. Validations of the simulation algorithm are performed for (i) terminal settling velocity of sedimenting spheroid in different orientations under the action of an external force like gravity, (ii) lateral drift of a settling spheroid with respect to the direction of gravity, (iii) in plane and out-of-plane (three-dimensional) rotations of the spheroid in a simple shear flow (Jeffery's orbits), and (iv) propulsion velocity of a spheroidal squirmer. This paper is organized as follows: In section~\ref{sec:method} we discuss the lattice Boltzmann simulation method, the dynamical description of the ellipsoids and the coupling scheme between the two algorithms. Results and discussions are contained in section~\ref{sec:results}, where the simulation method is validated and its numerical results are compared to available analytical solutions. Finally, in section~\ref{sec:conclusions} we conclude and discuss the scope of this work.

\section{Simulation Method}
\label{sec:method}
\begin{figure}
    \centering
    \includegraphics[trim = 100 50 0 50, clip, width=0.5\textwidth]{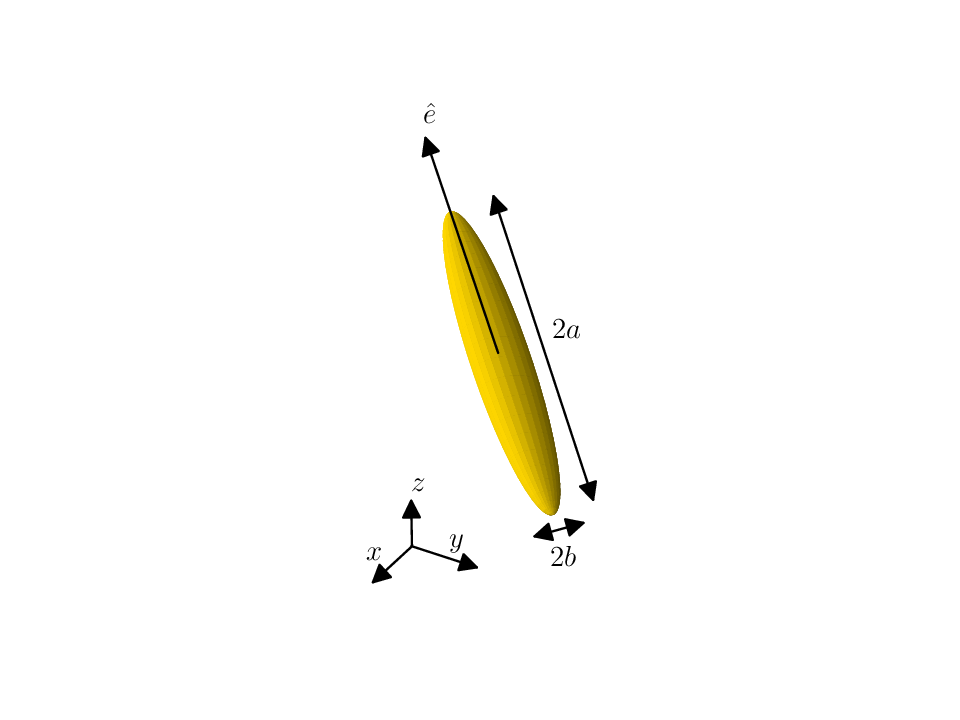}
    \caption{A prolate ellipsoid of length $2a$ (major axis) oriented along $\hat{e}$ with two minor axes $2b$ and $2c$ of equal length is considered in this work.}
    \label{fig:spheroid}
\end{figure}
We consider an ellipsoid of length $2a$ and minor axes of lengths $2b$ and $2c$ as shown in Fig.~\ref{fig:spheroid} suspended in a fluid. The instantaneous orientation of the major axis of the ellipsoid is indicated by a unit vector $\hat{e}$. Passive ellipsoids are apolar, and they have a head-tail symmetry, i.e. $\hat{e} \equiv -\hat{e}$ while active ellipsoids will have a polarity.

In the following we introduce the lattice Boltzmann method that describes the fluid dynamical aspects of the problem, the dynamical description of the ellipsoidal particles, and the boundary conditions that ensure the coupling between the two parts of the algorithm. The algorithm is implemented in the parallel lattice Boltzmann code \textit{Ludwig} \cite{Ludwig, desplat2001ludwig}.

\subsection{Lattice Boltzmann method}
As is common in the lattice Boltzmann literature \cite{kruger2017lattice}, the simulation domain is discretized using a geometric lattice with spacing $\Delta x = 1$, whereas the timestep size is $\Delta t = 1$. The simulations are performed using the $D3Q19$ and $D3Q27$ models to obtain further insights into how the results depend on the choice of numerical scheme. $D3$ represents the number of dimensions of the domain, while $Q19$ and $Q27$, respectively, represent 19 and 27 discrete lattice vectors $\bm{c}_i$ with $i = \{0, \dots, 18\}$ and $i = \{0, \dots, 26\}$, respectively (or equivalently the number of nearest neighbors) on the 3D cubic lattice. 

A discrete velocity distribution function $f_i(\bm{x},t)$ with velocity $\bm{c}_i$ is defined at each grid point $\bm{x}$ at time $t$ that travels along the $i^{th}$ discrete direction such that
\begin{align}
    \sum_i f_i (\bm{x},t) &= \rho(\bm{x},t) \\
     \sum_i f_i (\bm{x},t) \bm{c}_{i} &= \rho(\bm{x},t) \bm{u}(\bm{x},t),
\end{align}
where $\rho(\bm{x},t)$ and $\bm{u}(\bm{x},t)$ are the density and velocity of the fluid, respectively. This means the zeroth and first moment of the discrete distribution function $f_i$ with respect to the set of lattice directions $\{\bm{c}_i\}$ provide the hydrodynamic fields, namely the local density and momentum. 

The discrete distribution function $f_i$ undergoes collision and propagation steps consecutively. For a general collision operator ${\cal C}_i(t)$ this may be expressed
\begin{align}
f_i(\bm{x} + \bm{c}_i\Delta t, t + \Delta t) - f_i(\bm{x}, t) = \Delta t \, {\cal C}_i(t).
\end{align}
In its simplest form, the collision operator may be related to a single relaxation time $\tau$ as ${\cal C}_i(t) = [f_i(\bm{x}, t) - f_i^{eq}(\bm{x}, t)]/\tau$, where $f_i^{eq}$ is
the equilibrium distribution. The relaxation time determines the kinematic viscosity $\nu$ of the fluid via the relation $\nu = (\tau - \frac{1}{2})/3$ in the chosen convention \spt{\cite{kruger2017lattice}}. The form of the equilibrium distribution function, $f_i^{eq}$ is constrained by the governing equations to be recovered, namely the Navier Stokes equations:
\begin{align}
    f_i^{eq} = \rho w^{\bm{c}_i} \left( 1 + \frac{\bm{u}\cdot\bm{c}_i}{c_s^2} + \frac{(\bm{u}\cdot\bm{c}_i)^2}{2c_s^4} - \frac{\bm{u}\cdot\bm{u}}{2 c_s^2} \right)
\end{align}
where $c_s$ is the speed of sound and $w^{\bm{c}_i}$ are the weights associated with the chosen velocity set $\bm{c}_i$ of the $D3Q19$ or $D3Q27$ lattice Boltzmann model. The actual collision scheme in this work uses a single viscosity (relaxation time) for the non-conserved hydrodynamic modes, and an instantaneous relaxation for non-hydrodynamic modes (cf. \cite{nguyen2002lubrication}).

The macroscopic behavior of the fluid following the above described discrete algorithm is given by the Navier-Stokes equations:
\begin{align}
\nabla\cdot\bm{u} &= 0\\
\rho\left(\partial_t\bm{u}+\bm{u}\cdot\nabla\bm{u}\right)&=-\nabla P+\eta\nabla^2\bm{u}
\end{align}
in the limit of an incompressible fluid. Here, $P$ is the pressure field and $\eta$ is the dynamic viscosity of the fluid $\eta = \rho\nu$. \spt{The speed of sound $c_s$ is related to the fluid pressure $P$ and fluid density $\rho$ as $P = \rho c_s^2$, indicating the slightly compressible nature of the numerical scheme ~\cite{kruger2017lattice}.}

\subsection{Boundary conditions}
\label{sec:boundary conditions}
Consider an ellipsoid that has a translational velocity $\bm{U}$, an angular velocity $\bm{\varOmega}$, and a center of mass that is located at $\bm{x}_c$. The surface of the ellipsoid is defined through the quadratic relation
\begin{align}
(\bm{x}-\bm{x}_c)^T\bm{A}(\bm{x}-\bm{x}_c) = 1,
\label{eq:ellipsoidsurface}
\end{align}
where $\bm{x}$ is any point on the surface of the ellipsoid and $\bm{x}^T$ is the transpose of $\bm{x}$. The eigenvalues of the positive definite $3 \times 3$ matrix $\bm{A}$ are related to the inverse of the semi-axes of the ellipsoid via ($1/a^2, 1/b^2, 1/c^2$), while the normalized eigenvectors represent the axes with $\hat{e}$ being the principal axis, and specify therefore the orientation of the ellipsoid. \spt{A simple way to understand the matrix $\mathbf{A}$ is to consider an ellipsoid that is oriented along the coordinate axes, in which case $\mathbf{A}$ reduces to a diagonal matrix, and the equation of the ellipsoid is simply $\frac{x^2}{a^2}+\frac{y^2}{b^2}+\frac{z^2}{c^2} = 1$ \cite{strang2022introduction}.} The surface defined by Eq.~\ref{eq:ellipsoidsurface} intersects the links that connect the lattice nodes (the discrete points $\bm{x}$ in the domain) in the fluid and in the solid. Boundary nodes are defined to be halfway along the links, $\bm{x}_b = \bm{x} + \frac{1}{2}\bm{c}_b\Delta t$, and thus the boundary nodes represent an approximation to the surface of the ellipsoid. 

During the streaming-collision operations in the lattice Boltzmann algorithm, the populations $f_b$ located on the fluid node at $\bm{x}$ and connecting to a boundary node $\bm{x}_b$ via the lattice velocity vector $\bm{c}_b$ follow a half-way bounce-back scheme, which is known as bounce-back on the links \cite{ladd1994numerical}. There is no fluid inside the volume enclosed by the boundary nodes \cite{nguyen2002lubrication}. The bounce back of the fluid populations $f_b$ results in an exchange of momentum between the fluid and the solid nodes. The corresponding force exerted by the fluid on the solid per link may be calculated \cite{nguyen2002lubrication} as
\begin{align}
    \bm{F}_b(\bm{x}_b,t+\frac{1}{2}\Delta t) = \frac{\Delta x^3}{\Delta t} \left[ 2f_b^*(\bm{x},t) - \frac{2 w^{\bm{c}_b} \rho_0 \bm{u}_b\cdot\bm{c}_b}{c_s^2}\right] \bm{c}_b,
    \label{eq:fb}
\end{align}
where $f_b^*(\bm{x},t)$ is the post-collision distribution function, $\rho_0$ is the mean density of the fluid, and $\bm{u}_b$ is the local velocity of the boundary node calculated as 
\begin{align}
\bm{u}_b = \bm{U} + \bm{\varOmega} \times (\bm{x}_b - \bm{x}_c).
 \label{eq:ub}
\end{align}
The contributions $\bm{F}_b(\bm{x}_b,t+\frac{1}{2}\Delta t)$ from all boundary nodes are accumulated to obtain the total force on the ellipsoid $\bm{F}=\sum_b \bm{F}_b$. Similarly, the torque on the ellipsoid, $\bm{T}$ is obtained by adding the contributions $(\bm{x}_b - \bm{x}_c) \times \bm{F}_b$ from all boundary nodes. This procedure also ensures that the combined momentum of the fluid and the particle is conserved during the simulation.

\subsection{Dynamics of the ellipsoid}
In the formulation discussed above, the translational and angular velocities $\bm{U}$ and $\bm{\varOmega}$ of the ellipsoidal particle are to be calculated as part of the algorithm. An explicit update is usually unstable, and thus, following the procedure suggested by \citet{nguyen2002lubrication} for spherical particles, an implicit numerical evaluation scheme is proposed to update translational and angular velocities of the ellipsoidal particles. To facilitate this, the total force $\bm{F}$ and the torque $\bm{T}$ on the particle are written as
\begin{align}
\bm{F} &= \bm{F}_0 - \boldsymbol{\zeta}^{FU} \cdot \bm{U} - \boldsymbol{\zeta}^{F\varOmega} \cdot \boldsymbol{\varOmega}\label{eq:Fsplit}\\
\bm{T} &= \bm{T}_0 - \boldsymbol{\zeta}^{TU} \cdot \bm{U} - \boldsymbol{\zeta}^{T\varOmega} \cdot \boldsymbol{\varOmega}\label{eq:Tsplit}
\end{align}
where $\bm{F}_0$ and $\bm{T}_0$ are `velocity-independent' forces and torques, evaluated solely from the post-collision distributions:
\begin{align}
\bm{F}_0(t + \frac{1}{2}\Delta t) &= \frac{\Delta x^3}{\Delta t} \sum_b 2 f_b^*(\bm{x},t) \bm{c}_b,\label{eq:F0}\\
\bm{T}_0(t + \frac{1}{2}\Delta t) &= \frac{\Delta x^3}{\Delta t} \sum_b 2 f_b^*(\bm{x},t) (\bm{x}_b - \bm{x}_c) \times \bm{c}_b.\label{eq:T0}
\end{align}
The drag coefficient matrices in Eqs.~\ref{eq:Fsplit} and \ref{eq:Tsplit} may be calculated \cite{nguyen2002lubrication} as
\begin{align}
\boldsymbol{\zeta}^{FU} &= \frac{2 \rho_0 \Delta x^3}{c_s^2 \Delta t} \sum_b w^{\bm{c}_b} \bm{c}_b \bm{c}_b\label{eq:drag1}\\
\boldsymbol{\zeta}^{F\varOmega} &= \frac{2 \rho_0 \Delta x^3}{c_s^2 \Delta t} \sum_b w^{\bm{c}_b} \bm{c}_b (\bm{x}_b - \bm{x}_c) \times \bm{c}_b\\
\boldsymbol{\zeta}^{TU} &= \frac{2 \rho_0 \Delta x^3}{c_s^2 \Delta t} \sum_b w^{\bm{c}_b} ((\bm{x}_b - \bm{x}_c)\times\bm{c}_b) \bm{c}_b\\
\boldsymbol{\zeta}^{T\varOmega} &= \frac{2 \rho_0 \Delta x^3}{c_s^2 \Delta t} \sum_b w^{\bm{c}_b} ((\bm{x}_b - \bm{x}_c)\times\bm{c}_b)  ((\bm{x}_b - \bm{x}_c)\times\bm{c}_b). 
\label{eq:drag4}
\end{align}
Assuming that the drag coefficients are independent of time, discretized conservation equations of linear and angular momentum along with Eqs.~\ref{eq:Fsplit} and \ref{eq:Tsplit} are written down as
\begin{align}
&M \frac{\bm{U}(t+\Delta t) - \bm{U}(t)}{\Delta t} = \bm{F}_0 (t + \frac{1}{2} \Delta t) \nonumber\\ &\hspace{50pt}- \boldsymbol{\zeta}^{FU} \cdot \bm{U}(t + \Delta t) - \boldsymbol{\zeta}^{F\varOmega} \cdot \boldsymbol{\varOmega}(t + \Delta t) \label{eq:Fdiscrete}\\
&\bm{I}(t) \cdot \frac{\boldsymbol{\varOmega}(t+\Delta t) - \boldsymbol{\varOmega}(t)}{\Delta t} + \frac{d\bm{I}}{dt} \cdot  \boldsymbol{\varOmega}(t+\Delta t) \nonumber\\&= \bm{T}_0 (t + \frac{1}{2} \Delta t) - \boldsymbol{\zeta}^{TU} \cdot \bm{U}(t + \Delta t) - \boldsymbol{\zeta}^{T\varOmega} \cdot \boldsymbol{\varOmega}(t + \Delta t) \label{eq:Tdiscrete}
\end{align}
where $M$ and $\bm{I}$ are the mass and moment of inertia tensor of the ellipsoidal particle respectively. The time dependence of $I(t)$ will be discussed in detail in the following section. Hence, we obtain two linear equations in $\bm{U}(t+\Delta t)$ and $\boldsymbol{\varOmega}(t+\Delta t)$ from Eqs.~\ref{eq:Fdiscrete} and \ref{eq:Tdiscrete}, which are solved using the Gaussian elimination method. Since the formulation is implicit, the stability of the algorithm is guaranteed as well.

Thus, solving the linear system consisting of Eqs.~\ref{eq:Fdiscrete} and \ref{eq:Tdiscrete} determines the unknowns $\bm{U}(t+\Delta t)$ and $\boldsymbol{\varOmega}(t+\Delta t)$. The only remaining unknowns are the new position and orientation of the particle. The position of the particle is simply evaluated as
\begin{align}
\bm{x}_c(t + \Delta t) = \bm{x}_c (t) + \frac{1}{2}\left( \bm{U}(t + \Delta t) + \bm{U}(t) \right).
\label{eq:posupdate}
\end{align}
Similarly, the mean angular velocity $
\tilde{\boldsymbol{\varOmega}}(t) = \frac{1}{2} \left(\boldsymbol{\varOmega}(t) + \boldsymbol{\varOmega}(t + \Delta t) \right)$ is used to update the orientation of the particles. However, unlike in case of spherical particles, the orientation dynamics of ellipsoids, namely the time evolution of its principal axes, requires careful consideration. Following  \citet{goldstein2011classical} the three Euler angles $\phi, \theta, \psi$ are defined in the co-moving frame of the particle. $\phi$ is the rotation angle around the $z$-axis, whereas $\theta$ and $\psi$ are the rotation angles around the resulting intermediate $x'$- and $z'$-axis, respectively. However, to avoid accumulated errors from successive matrix operations and singular matrix operations, it is convenient to introduce unit quaternions \cite{voth2017anisotropic, zhao2013direct, zhao2013novel, fan1995sublayer, fan2000wall, van2015modelling} (also referred to as Euler parameters) defined as
\begin{align}    \mathbf{q}&\equiv[\cos{(\theta/2)}\cos{((\phi+\psi)/2)},\sin{(\theta/2)}\cos{((\phi-\psi)/2)},\nonumber\\&\sin{(\theta/2)}\sin{((\phi-\psi)/2)},\cos{(\theta/2)}\sin{((\phi+\psi)/2)}],
\end{align}
The temporal evolution of $\mathbf{q}$ determines the instantaneous Euler angles and thus the orientation of the ellipsoid.

Knowing the angular velocity of the ellipsoid $\boldsymbol{\Omega}$, the evolution equation for the quaternion may be written as
\begin{align}
    \Dot{\mathbf{q}} = \frac{1}{2}\mathbf{\Omega}\,\mathbf{q},
    \label{eq:qevolution}
\end{align}
where $\mathbf{\Omega}=(0, \bm{\varOmega})$ is a pure quaternion with vectorial part $\bm{\varOmega}$.
However, following \cite{zhao2013direct, zhao2013novel, van2015modelling}, we suggest the procedure given below to determine the time evolution of the quaternions,
\begin{align}
&\tilde{\mathbf{q}} (t) = \left[\cos{\frac{||\tilde{\boldsymbol{\varOmega}} (t)||\Delta t}{2}},\sin{\frac{||\tilde{\boldsymbol{\varOmega}} (t))||\Delta t}{2}}\frac{\tilde{\boldsymbol{\varOmega}}(t))}{||\tilde{\boldsymbol{\varOmega}}(t))||}\right]\label{eq:qnplus0}\\
&\mathbf{q}(t+\Delta t) = \tilde{\mathbf{q}}(t)\mathbf{q}(t) ,
\label{eq:qnplus1}
\end{align}
thus avoiding numerical integration of Eq.~\ref{eq:qevolution}. In the above, $||\cdot||$ indicates the norm of the vector. The magnitude of the quaternion is exactly unity and therefore, this procedure avoids any requirement of renormalization to account for errors from numerical integration. Thus, both position (Eq.\ref{eq:posupdate}) and orientation (Eq.~\ref{eq:qnplus1}) are updated based on the mean value of translational and angular velocity at $t$ and $t + \Delta t$.

\subsection{Moment of inertia of the ellipsoid}
The left hand side of Eq.~\ref{eq:Tdiscrete} is based on the Euler equations for rigid body rotation,
\begin{align}
\bm{T} = \frac{d}{dt}(\bm{I}\cdot\boldsymbol{\varOmega}) = \bm{I}\cdot\frac{d\boldsymbol{\varOmega}}{dt} + \frac{d\bm{I}}{dt}\cdot\boldsymbol{\varOmega}.
\label{eq:Tdiff}
\end{align}
The second term in Eq.~\ref{eq:Tdiff} vanishes for spherical particles as their moment of inertia tensor is constant in the laboratory frame. This is not the case for ellipsoidal particles, and the rate of change of the moments of inertia requires consideration. In the literature on molecular dynamics and other particle-based simulation techniques, Eq.~\ref{eq:Tdiff} is usually written in a body-fixed coordinate system so that the second term in Eq.~\ref{eq:Tdiff} vanishes even for anisotropic shapes. However, we avoid this route as for algorithmic purposes it is more appropriate to formulate the implicit solution method for updating $\bm{U}$ and $\boldsymbol{\varOmega}$ in the laboratory frame following Eqs.~\ref{eq:Fdiscrete} and \ref{eq:Tdiscrete}. Below we discuss the time dependence of the moment of inertia tensor $\bm{I}(t)$ and its time derivative that may be used in evaluating terms in Eq.~\ref{eq:Tdiff}.

In the body frame of reference defined by the principal axes of the ellipsoid, the moment of inertia tensor is
\begin{align}
\bm{I} &=
\begin{bmatrix}
\frac{1}{5} M(b^2+c^2) & 0 & 0 \\
0 & \frac{1}{5} M (a^2 + c^2) & 0 \\
0 & 0 & \frac{1}{5} M(a^2+b^2)
\end{bmatrix}\\
&=
\begin{bmatrix}
I_1 & \quad 0 & \quad 0 \\
0 & \quad I_2 & \quad 0 \\
0 & \quad 0 & \quad I_3
\end{bmatrix}
\end{align}
Since the ellipsoid undergoes rotational motion, but the angular momentum equation (Eq.~\ref{eq:Tdiscrete}) is given in the laboratory frame, the moment of inertia tensor $\bm{I}$ has to be determined at every time step. This may be done using the unit quaternion $\mathbf{q}(t)$ as follows. Any pure quaternion $\mathbf{s}=(0,\bm{s})$ with vectorial part $\bm{s}=(s_1, s_2, s_3)$ in the body frame can be rotated using the unit quaternion $\mathbf{q} = (q_0, \bm{q})$ with vectorial part $\bm{q}=(q_1,q_2,q_3)$ and its inverse $\mathbf{q}^{-1}=(q_0, -\bm{q})$ \cite{rapaport2004art, zhao2013novel}:
\begin{align}
\mathbf{s}' = \mathbf{q}\,\mathbf{s}\,\mathbf{q}^{-1}
\label{eq:vectrotn}
\end{align}
The quaternion algebra provides the explicit expression for the vectorial part of $\mathbf{s}'$: $\bm{s}' = (2q_0^2-1)\bm{s} + 2(\bm{q}\cdot\mathbf{s})\bm{q} + 2q_0 \bm{q} \times \bm{s}$. Repeated application of Eq.~\ref{eq:vectrotn} gives the moment of inertia of the rotated ellipsoid \cite{fan1995sublayer, zhao2013direct, van2015modelling}, 
\begin{align}
\mathbf{I}' = \left(\mathbf{q} \left(\mathbf{q}\,\mathbf{I}\,\mathbf{q}^{-1}\right)^T\mathbf{q}^{-1}\right)^T ,
\label{eq:Irot}
\end{align}
where $^T$ is the matrix transpose and $\mathbf{I}'$ and $\mathbf{I}$ are pure quaternion moment of inertia tensors, so for instance
\begin{align}
\mathbf{I} &=
\begin{bmatrix}
0 & \quad 0   & \quad 0   & \quad 0\\
0 & \quad I_1 & \quad 0   & \quad 0\\
0 & \quad 0   & \quad I_2 & \quad 0\\
0 & \quad 0   & \quad 0   & \quad I_3
\end{bmatrix}.
\end{align} 
Therefore, the time-dependent moment of inertia tensor $\bm{I}(t)$ can be easily determined from the quaternion $\mathbf{q}(t)$.

The first term in Eq.~\ref{eq:Tdiff} assumes that the moment of inertia tensor $\bm{I}(t)$ is determined at time $t$ and not at $t + \Delta t$. 
This procedure can be improved with a predictor-corrector method which determines $\bm{I}(t + \Delta t)$, but it is not pursed in this work.

There are two possibilities to calculate the time derivative of the moment of inertia tensor in the second term in Eq.~\ref{eq:Tdiscrete}. The first method is to adopt a simple, finite difference  approximation,
\begin{align}
\frac{d\bm{I}}{dt} = \frac{\bm{I}(t) -\bm{ I}(t - \Delta t)}{\Delta t}.
\end{align}
The second, and more elegant, route is to avoid numerical differentiation of the temporal derivative, but differentiate Eq.~\ref{eq:Irot} directly. With repeated application of product rule of differentiation of Eq.~\ref{eq:Irot} we have,
\begin{align}
\frac{d\mathbf{I}}{dt} &= \left(\dot{\mathbf{q}} \left(\mathbf{q}\,\mathbf{I}\,\mathbf{q}^{-1}\right)^T\mathbf{q}^{-1}\right)^T + \left(\mathbf{q} \left(\dot{\mathbf{q}}\,\mathbf{I}\,\mathbf{q}^{-1}\right)^T\mathbf{q}^{-1}\right)^T \nonumber\\ 
&+ \left(\mathbf{q} \left(\mathbf{q}\,\mathbf{I}\,\dot{\mathbf{q}}^{-1}\right)^T\mathbf{q}^{-1}\right)^T + \left(\mathbf{q} \left(\mathbf{q}\,\mathbf{I}\,\mathbf{q}^{-1}\right)^T\dot{\mathbf{q}}^{-1}\right)^T.
\label{eq:Irotdiff}
\end{align}
The time derivative of the quaternion $\dot{\mathbf{q}}$ may be determined using Eq.~\ref{eq:qevolution}, and hence
each term in Eq.~\ref{eq:Irotdiff} may be easily evaluated. Thus, this second method to evaluate $d\bm{I}/dt$ avoids numerical differentiation altogether. In our simulations, no discernible difference in the stability of the simulation was observed between the two methods mentioned above.

\spt{The complete algorithm can be summarized as follows: The streaming and collision step are performed in the usual way. The two-way coupling of fluid-structure interaction is realized on one side during the streaming step, when mid-grid bounce back conditions are applied on those populations which stream to a solid node, and on the other side through the solid nodes, which are determined at every time step from the updated position and orientation of the ellipsoid using the above implicit method. All operations are carried out in the laboratory frame of reference. The use of quaternion algebra make the calculation of the angular velocity, moment of inertia and orientation more straightforward. The individual steps of the algorithm are detailed in Appendix~\ref{appA}.}

\section{Results and Discussion}
\label{sec:results}

Four different cases were analyzed using the lattice Boltzmann formulation discussed in the previous section: (i) a sedimenting spheroid in different orientations, (ii) drifting of a sedimenting spheroid in an inclined orientation, (iii) kinematics of a spheroid in a simple shear flow, and (iv) dynamics of an active, spheroidal microswimmer. The results from each case are discussed below. \spt{All simulations corresponding to case (i), (ii) and (iv) have been carried in a cubic domain with periodic boundary conditions. In case (iii), the spheroid is placed between two moving parallel plates. All simulations are initialized with a quiescent fluid in the domain. The simulations are performed until a steady state is reached and the obtained results are discussed below.}

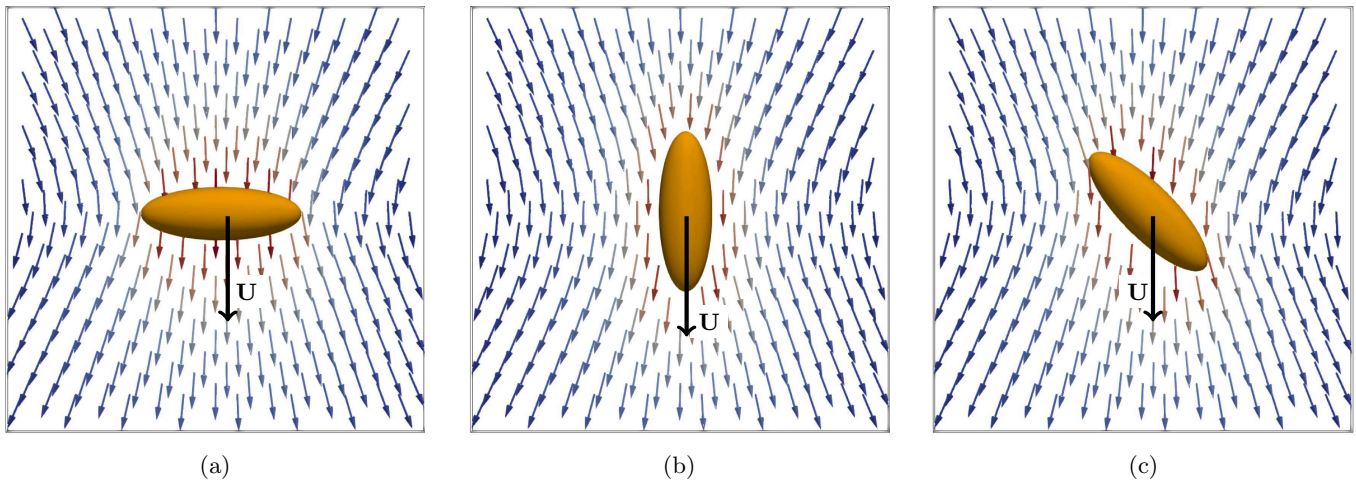
\begin{figure*}[ht!]
     \begin{tikzpicture}
    \node (image) at (0,0) {
          \centering
\begin{subfigure}[b]{0.32\textwidth}
         \centering
         \includegraphics[trim = 0 0 60 220, clip, width=\textwidth, angle=-90]{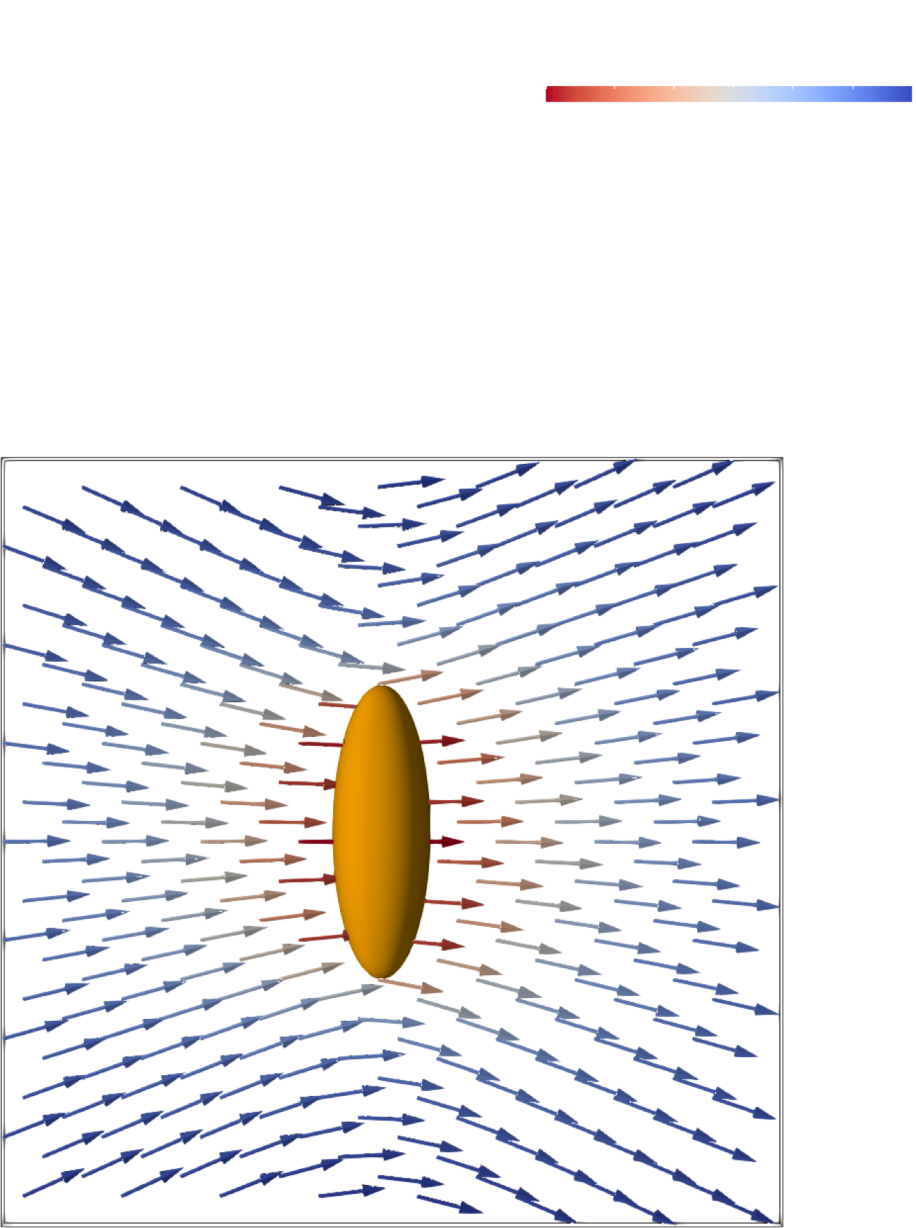}
         \caption{}
     \end{subfigure}
     \quad
     \hfill
     \begin{subfigure}[b]{0.32\textwidth}
         \centering
         \includegraphics[trim = 0 0 60 220, clip, width=\textwidth, angle=-90]{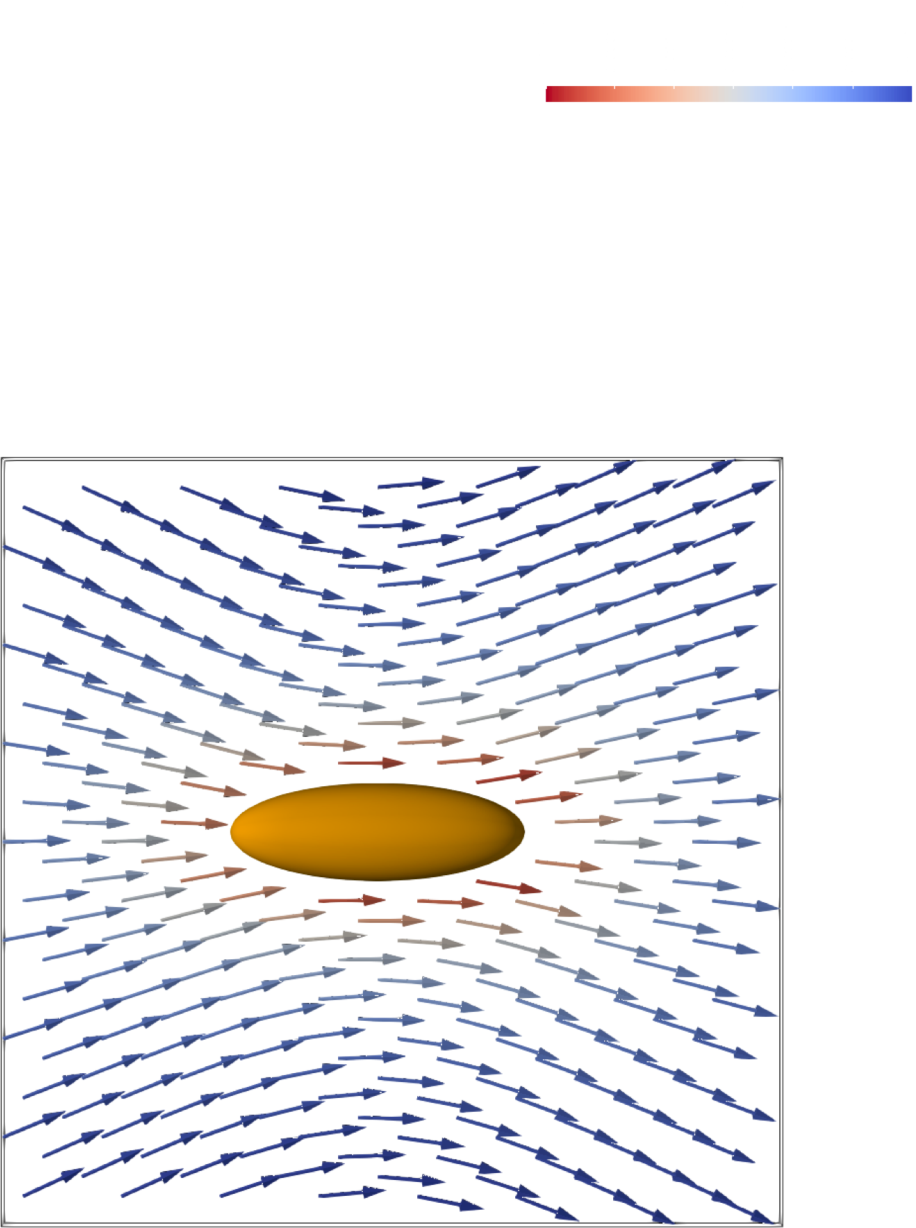}
         \caption{}
     \end{subfigure}
     \quad
    \hfill
     \begin{subfigure}[b]{0.32\textwidth}
         \centering
         \includegraphics[trim = 0 165 60 190, clip, width=\textwidth, angle=-90]{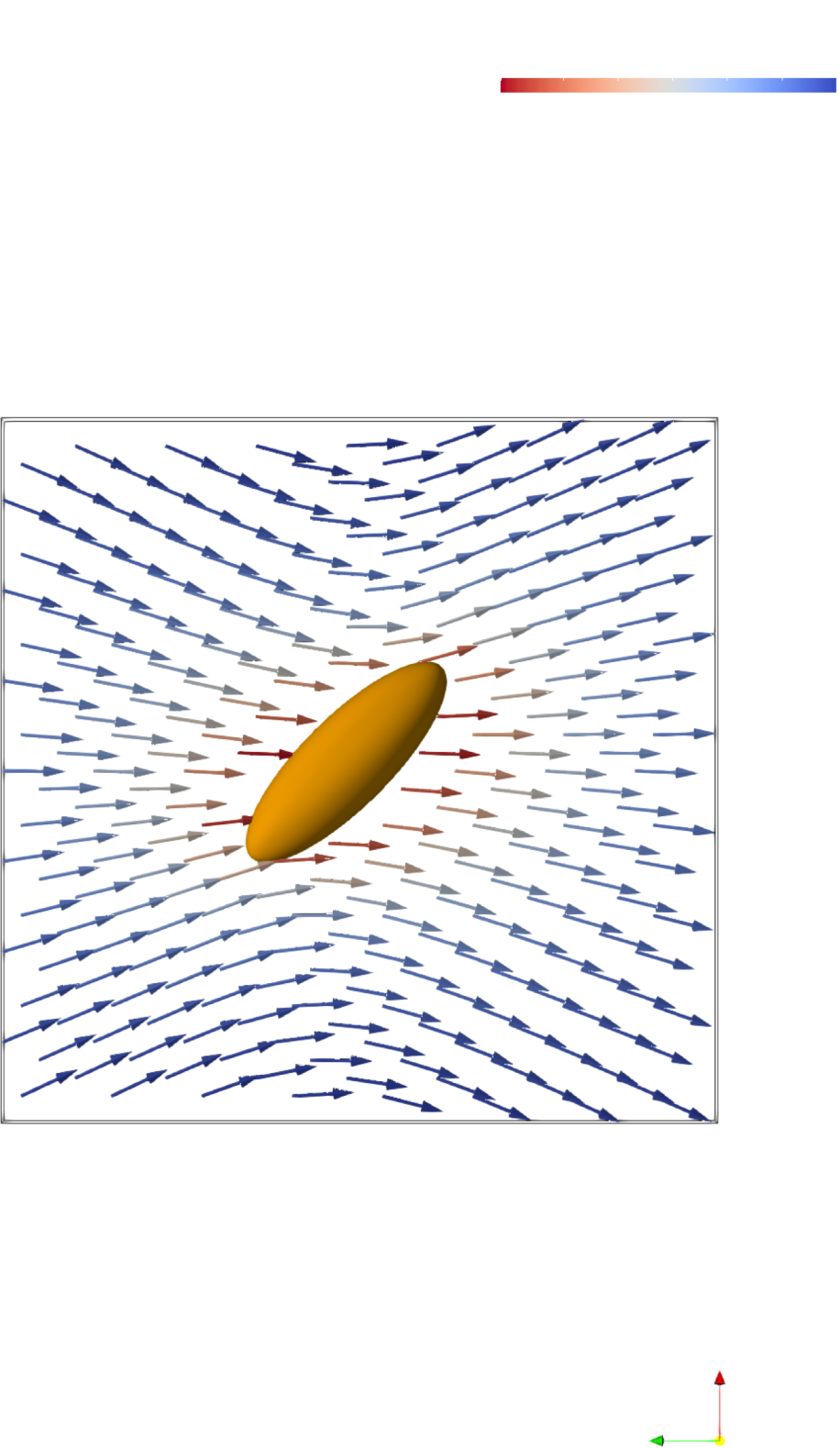}
         \caption{}
     \end{subfigure}
    };
    \node[fill=white] at (-5.75,-0.6) {$\mathbf{U}$};
    \draw[->,ultra thick](-6,0.4) -- (-6,-1.0);
    \node[fill=white] at (0.4,-1) {$\mathbf{U}$};
    \draw[->,ultra thick](0.1,0.4) -- (0.1,-1.2);
    \node[fill=white] at (6.1,-0.6) {$\mathbf{U}$};
    \draw[->,ultra thick](6.3,0.4) -- (6.3,-1.0);
\end{tikzpicture}
        \caption{Fluid velocity in the laboratory frame developed around a settling spheroid ($a > b = c$) in three different orientations, the long axis of the spheroid oriented (a) perpendicular to the direction of gravity, $\hat{e} \perp \hat{g}$, (b) parallel to the direction of gravity, $\hat{e} \parallel \hat{g}$ and (c) inclined at an angle $45^{\circ}$ to the direction of gravity. The black arrow in each figure indicates the velocity of the spheroid. \spt{The fluid velocity vectors are colored using a `jet' color map such that red arrows indicate larger velocity than blue arrows.}}
        \label{fig:velsett}
\end{figure*}

\subsection{Sedimenting spheroid}
\label{sec:sedimenting spheroid}

Consider a spheroid with orientation vector $\hat{e}$ as shown in Fig.~\ref{fig:spheroid} sedimenting under the action of an external force, say gravity in an otherwise quiescent fluid. Let $\hat{g}$ be the unit vector indicating the direction of the external force. Under the action of the external force, the spheroidal particle accelerates, but the fluid drag opposes this motion. Balancing the two opposing forces, the particle undergoes a steady translation with a constant velocity, usually referred to as terminal settling velocity in the context of gravitational sedimentation.

Fig.~\ref{fig:velsett} shows the fluid flow developed around the sedimenting spheroid at its terminal settling velocity in three different orientations (a) when the long axis of the spheroid is perpendicular to the direction of the external force, $\hat{e} \perp \hat{g}$ (broad-side on) (b) when the long axis is parallel to the direction of the external force, $\hat{e} \parallel \hat{g}$ (end-on), and (c) when the long axis forms a $45^{\circ}$ angle with the direction of gravity. The arrows in the figure indicate the direction of the velocity vectors in the fluid in the laboratory frame of reference. The color code gives the magnitude of the velocity, where blue and red mark the lowest and highest velocities, respectively. As expected, the fluid velocity has a maximum at the surface of the translating spheroid, and it decays further away from the particle. The flow fields are similar in all cases. There is also a qualitative similarity to the well-known Stokeslet flow. In the broad-side on orientation, the flow field appears laterally extended compared to that of a sphere. In the end-on orientation of the spheroid, the flow appears extended in the vertical direction, in consonance with the geometry of the sedimenting particle. Fig.~\ref{fig:velsett}(c) is discussed in the next subsection.

\begin{figure*}[htbp!]
     \centering
     \begin{subfigure}[b]{0.48\textwidth}
         \centering
         \includegraphics[width=\textwidth]{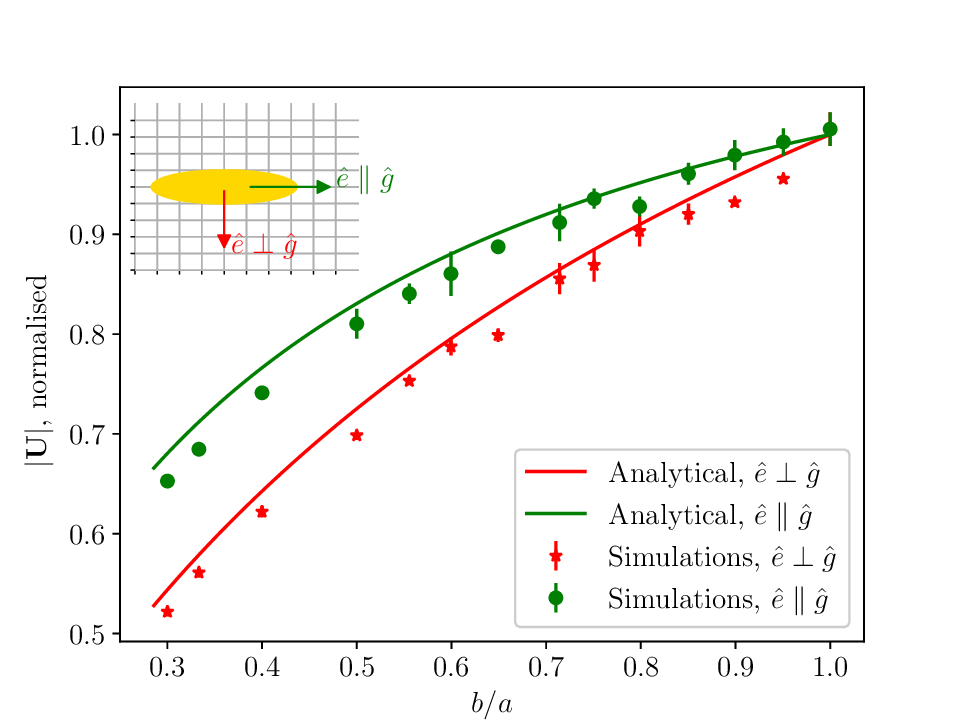}
         \caption{}
     \end{subfigure}
     \hfill
     \begin{subfigure}[b]{0.48\textwidth}
         \centering
         \includegraphics[width=\textwidth]{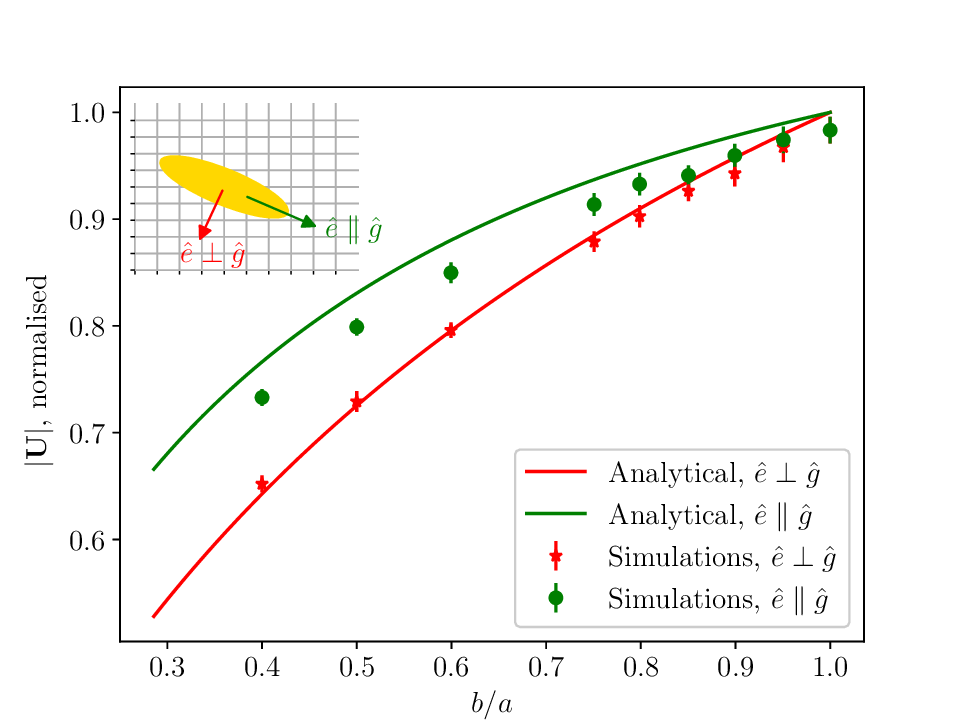}
         \caption{}
     \end{subfigure}
     \hfill
     \begin{subfigure}[b]{0.48\textwidth}
         \centering
         \includegraphics[width=\textwidth]{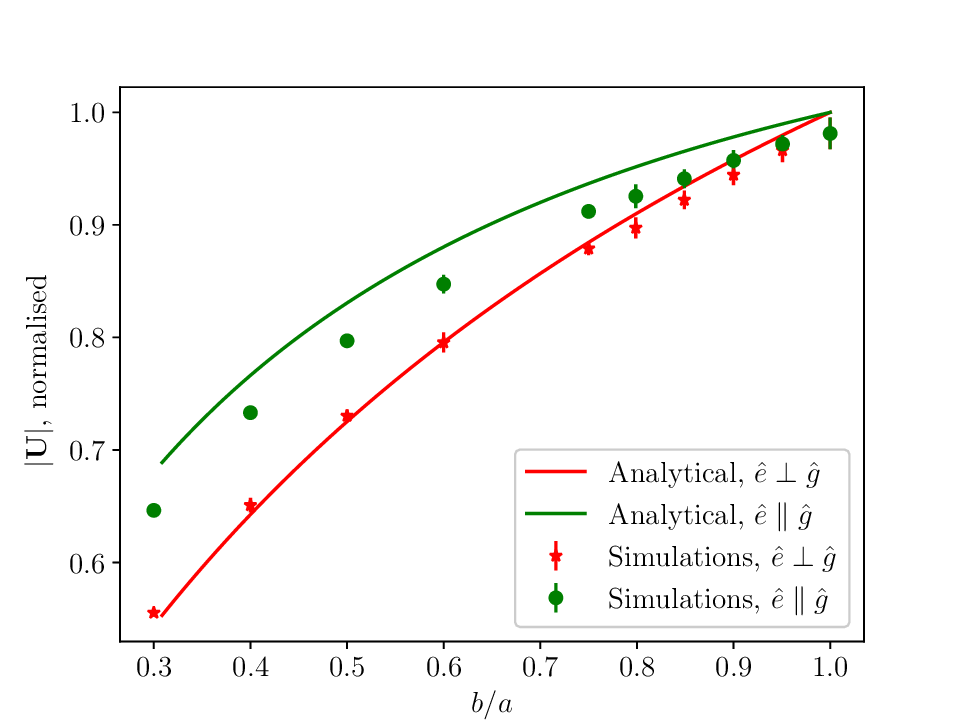}
         \caption{}
     \end{subfigure}
     \hfill
     \begin{subfigure}[b]{0.48\textwidth}
         \centering
         \includegraphics[width=\textwidth]{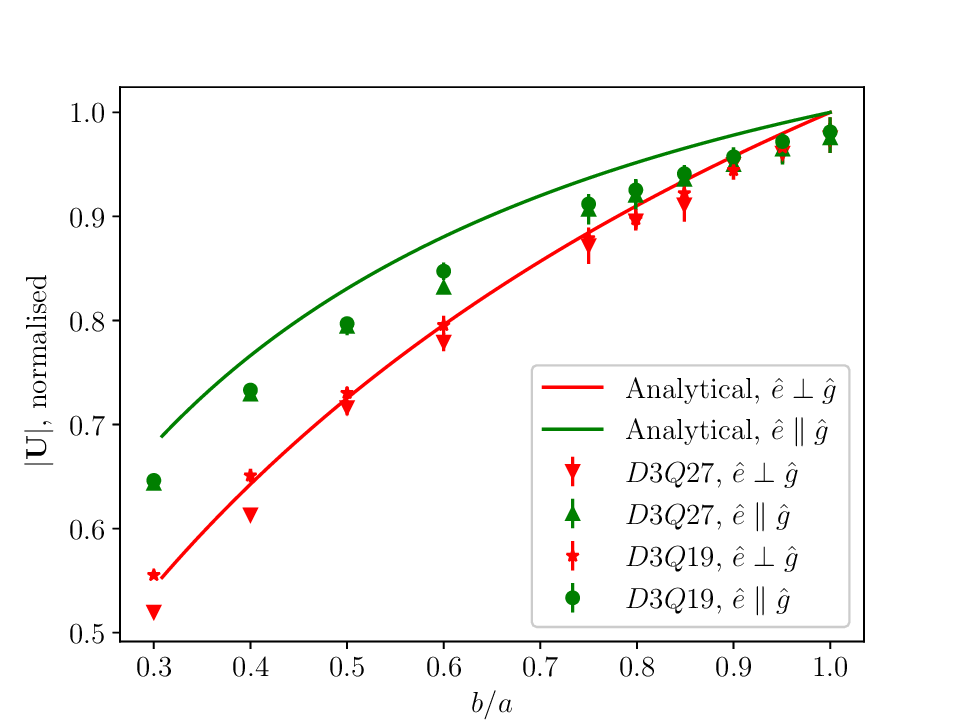}
         \caption{}
     \end{subfigure}
        \caption{Settling velocity of a spheroid ($a > b = c$) in two different orientations, the long axis of the spheroid oriented (i) perpendicular to the direction of gravity, $\hat{e} \perp \hat{g}$ (red line, $\vcenter{\hbox{\Large{$\bullet$}}}$) and (ii) parallel to the direction of gravity, $\hat{e} \parallel \hat{g}$ (green line, $\scriptstyle\blacksquare$). Simulations of spheroids (a) with $b = 2.5$, $\nu = 0.1$, and when $\hat{e}$ are  $\hat{g}$ are along the principal directions of the grid (as shown in the inset), (b) with $b = 2.5$, $\nu = 0.1$ but $\hat{e}$ are  $\hat{g}$ are oriented at $30^{\circ}$ to the principal directions of the grid (as shown in the inset) (c) with  LB parameters as reported in \cite{nguyen2002lubrication} namely $b = 2.7$, $\nu = 0.1667$ for spherical particles (see text). (d) comparison using $D3Q19$ and $D3Q27$ LBM models with $b = 2.7$, $\nu = 0.1667$. In all cases, simulations have been done in a domain of $256^3$. The reported settling velocity $|\bm{U}|$ is normalized to the settling velocity of a sphere with radius $r=b$. The error bars indicate the error in the calculations originating from the discrete shape of the particle, and are smaller than the symbol in most cases.}
        \label{fig:settling}
\end{figure*}

At zero Reynolds number, the terminal settling velocity of the spheroid in the two orientations can be exactly determined. If $\bm{F}_e$ is the external force acting on the spheroid, then the translational velocity can be determined as \cite{leal2007advanced}
\begin{align}
\bm{F}_e &= -6 \pi \mu a (U_{\parallel} C_{\parallel} \hat{e} + U_{\perp} C_{\perp} \hat{e}_{\perp})\label{eq:settlingvel}\\
C_{\parallel} &= \frac{8}{3} \epsilon^3 \left[ -2\epsilon + (1+\epsilon^2) \log\frac{1+\epsilon}{1-\epsilon} \right]^{-1}\\
C_{\perp} &= \frac{16}{3} \epsilon^3 \left[ 2\epsilon + (3\epsilon^2-1) \log\frac{1+\epsilon}{1-\epsilon} \right]^{-1}
\end{align}
where $\hat{e}_{\perp}$ is the direction normal to the long-axis and $\epsilon = \sqrt{1 - b^2/a^2}; 0 \le \epsilon < 1$ is the eccentricity of the spheroid.

The comparison between the sedimentation velocity of the spheroid obtained from simulations and that determined from the analytical expression (Eq.~\ref{eq:settlingvel}) is shown in Fig.~\ref{fig:settling}. In each subplot in this figure, the $x-$axis represents the aspect ratio of the spheroid determined as the ratio of length along the minor axis to length along the major axis, $b/a$; the $y-$ axis represents the terminal settling velocity of the spheroid normalized with that of a sphere of radius $b$. Parameters chosen in these simulations correspond to a Reynolds number $< 0.01$.

Fig.~\ref{fig:settling}(a) shows the results obtained for (i) a broad-side on sedimenting spheroid, \textit{i.e.,} when $\hat{e} \perp \hat{g}$ and an end-on sedimenting spheroid,  \textit{i.e.,} when $\hat{e} \parallel \hat{g}$ in an otherwise quiescent fluid. A schematic of the configurations is shown in the figure inset. The symbols are the data obtained from the simulation and the continuous curves are the analytical predictions based on Eq.~\ref{eq:settlingvel}. When $b/a = 1$, the simulations correspond to that of a settling sphere. As the aspect ratio of the spheroid increases ($b/a$ decreases) the terminal settling velocity of the spheroid decreases. In the simulations, the external force acting on the spheroid and the length of the minor axis $b$ are not changed. Therefore, increasing the aspect ratio corresponds to a larger drag force on the spheroid, and hence a reduction in the terminal settling velocity, irrespective of the orientation of the spheroid. Secondly, the spheroid in broad-side on orientation sediments at a slightly smaller terminal settling velocity than the spheroid in end-on orientation, and this difference increases with increase in aspect ratio. All these features predicted by the analytical expression (Eq.~\ref{eq:settlingvel}) are captured by the simulations.

Fig.~\ref{fig:settling}(a) has shown the results of the cases when the unit vector $\hat{e}$ and $\hat{g}$ are oriented along one of the coordinate directions of the discrete domain, as in the inset. To check the generality of the method developed, in particular, to test any dependency of the accuracy of the results when $\hat{e}$ and $\hat{g}$ are not aligned with the principal directions of the grid, sedimentation simulations are performed by orienting the spheroid at a finite angle to the coordinate directions, as shown in the inset in Fig.~\ref{fig:settling}(b). Again, both the broad-side on and the end-on orientation are tested. The obtained results are plotted in Fig.~\ref{fig:settling}(b) as a function of the aspect ratio. The reduction in the terminal settling velocity with increase in aspect ratio, the larger terminal settling velocity in the end-on orientation compared to broad-side on orientation, and increase in the difference in the terminal settling velocity in two orientations with increase in aspect ratio are all captured by the simulations. These results confirm the validity of the method implemented.

While the results presented in both Fig.~\ref{fig:settling}(a) and (b) show a reasonable match between the analytical predictions and lattice Boltzmann simulations, small differences between the two calculations may also be noticed. The spread of the data indicates that the error may be associated with numerical discretisation of the particle surface as discussed in section~\ref{sec:boundary conditions}. The implementation of the bounce back scheme on the surface of the ellipsoidal particle is only first order accurate in $\Delta x$, but previous work on simulating suspensions of spherical particles have shown that the error arising from a first order bounce back scheme can be considerably reduced by an appropriate choice of the fluid viscosity and radius of the spherical particle \cite{nguyen2002lubrication}. This is possible because (i) the error in the boundary conditions are dependent on the viscosity of the fluid and may alter the "hydrodynamic radius" at which no-slip boundary conditions exist, (ii) the variance in discrete shape of the particle compared to the spherical shape is not a monotonic function of the nominal radius, but it gives rise to certain "favorable radii" at which the variance is smaller. Therefore, we performed simulations of a sedimenting spheroid by selecting the kinematic viscosity of the fluid $\nu = 1/6$ and the length of the minor axis $b = 2.7$, one of the choices suggested for spherical particles in \cite{nguyen2002lubrication}. The results obtained for both broad-side on and end-on orientation are shown in Fig.~\ref{fig:settling}(c). No considerable improvement in the results is observed compared to Fig.~\ref{fig:settling}(a)-(b). This lack of improvement by choosing special values of kinematic viscosity and size indicates that such special choices are restricted solely to spherical particles;  non-spherical particles have a range of lengths associated with it (lengths varying between $a$ and $b$ for a spheroidal particle), and further improvement in the accuracy of the method can  be achieved by improving the spatial resolution of the simulations.

The final test case reported in Fig.~\ref{fig:settling}(d) is to compare the results of two different LB models: $D3Q19$ and $D3Q27$. The latter model consists of a larger number of discrete velocity directions, yet both models show comparatively similar results. Thus, increasing the number of discrete velocity directions from 19 to 27 did not result in any significant improvement on comparing with analytical predictions, but the similarity of results demonstrates the robustness of the proposed numerical algorithm irrespective of the choice of the lattice Boltzmann model.

\begin{figure}
    \centering
    \includegraphics[width=\linewidth]{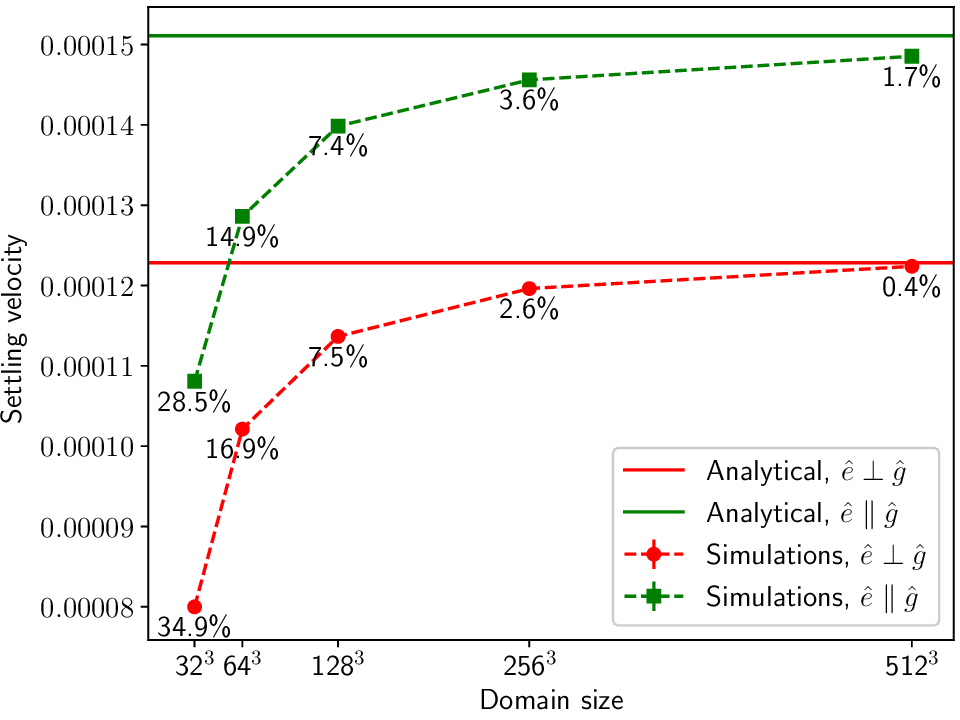}
    \caption{Effect of domain size on the accuracy of the settling velocity of a spheroid in two different orientations, the long axis of the spheroid oriented (i) perpendicular to the direction of gravity, $\hat{e} \perp \hat{g}$ (red line, $\vcenter{\hbox{\Large{$\bullet$}}}$) and (ii) parallel to the direction of gravity, $\hat{e} \parallel \hat{g}$ (green line, $\scriptstyle\blacksquare$). In the simulations $a = 7.5$, $b = 2.5$ and $\nu = 0.1$. The error calculated with respect to the analytical solution in each case is indicated below the marker.}
    \label{fig:settlingdomainsize}
\end{figure}

All the simulations reported in Fig.~\ref{fig:settling} are performed by imposing periodic boundary conditions on the domain. While this choice approximates a bulk fluid in the limit of large system size, the central particle inevitably interacts hydrodynamically with its images. Therefore, the simulation is equivalent to a periodic array of spheroids settling in the fluid. While analytical solutions are available to determine the settling velocity of an array of spheres \cite{hasimoto1959periodic}, no such approaches are available for settling ellipsoidal particles. Therefore, to isolate this effect due to periodic images and quantify the contribution from the hydrodynamic interactions, simulations were performed in cubic domains of size spanning from $32^3$ to $512^3$. Both broad-side on and end-on orientations were considered. The results are illustrated in Fig.~\ref{fig:settlingdomainsize} where the settling velocity is plotted as a function of the domain size. The horizontal lines in the figure indicate the analytical predictions according to Eq.~\ref{eq:settlingvel}. The numbers annotating each data point are the percentage difference in the result from the simulations compared to the analytical predictions. It can clearly be seen that in small domains hydrodynamic interactions with the periodic images result in large deviations ($>25\%$) from the analytical result. However, an increase in the size of the domain by an order of magnitude decreases the deviation in the settling velocity also by an order of magnitude. 

\subsection{Sedimentation of an inclined spheroid}
\label{sec:inclined}

\begin{figure*}
    \centering
    \begin{subfigure}[b]{0.45\textwidth}
            \includegraphics[width=1.1\linewidth]{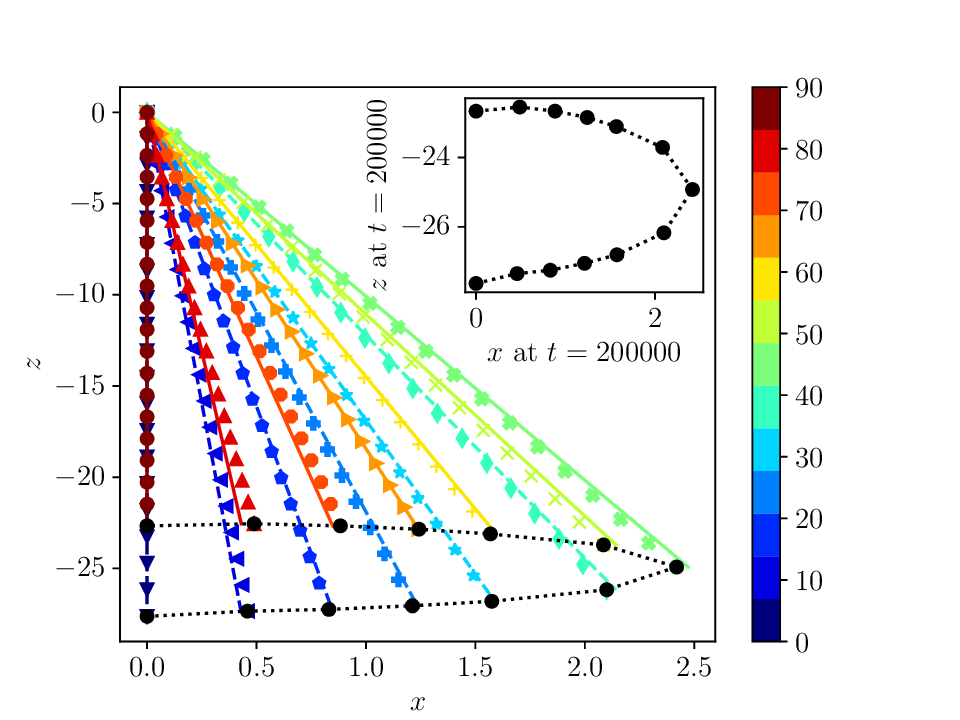}
            \caption{}
            \label{fig:inclined}
    \end{subfigure}
    \begin{subfigure}[b]{0.45\textwidth}
            \includegraphics[width=0.95\linewidth]{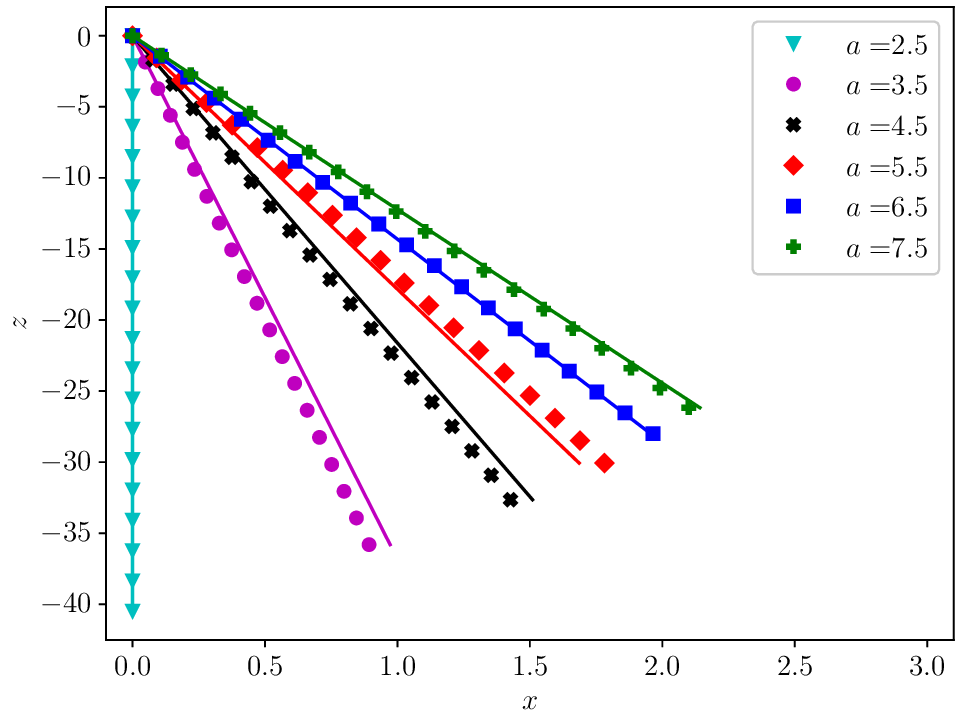}
            \caption{}
            \label{fig:inclined30}
    \end{subfigure}
    \caption{Trajectory of a spheroid settling at an angle to gravity. (a) $a = 7.5$ and $b = 2.5$. The angle of inclination, $\alpha$ is varied from $0$ to $90^{\circ}$ as indicated in the color bar. The initial position of the spheroid is $(0,0)$, and the final locations of the spheroid at the end of simulations ($t = 200000$ steps) are highlighted with a black circle ($\vcenter{\hbox{\Large{$\bullet$}}}$). The final locations ($(x,z)$) are also plotted in the inset. (b) The aspect ratio is varied by varying $a$ keeping $b = 2.5$. The continuous lines are from analytical predictions and the symbols are from the lattice Boltzmann simulations. In the analytical calculations for the trajectories reported in (a) and (b) the settling velocity in the two principal directions of the spheroid ($\hat{e} \perp \hat{g}$ and $\hat{e} \parallel \hat{g}$) are taken from the lattice Boltzmann simulations directly.}
\end{figure*}

In this section we consider a sedimenting spheroid when it is neither in broad-side on nor in end-on orientation, \textit{i.e.}, $\hat{e}$ and $\hat{g}$ are neither parallel nor perpendicular to each other. An example of the velocity field developed around the sedimenting spheroid, in an otherwise quiescent fluid, in this inclined orientation is shown in Fig.~\ref{fig:velsett}(c). Here, the direction of the external force $\hat{g}$ is acting downwards, but the spheroid is oriented at a $45^{\circ}$ angle to $\hat{g}$. Owing to the reversibility constraints imposed by the low Reynolds number hydrodynamics, the sedimenting spheroid retains its orientation during the simulation and reaches a terminal settling velocity. 

Even though the sedimenting spheroid does not rotate \spt{\cite{guazzelli2011physical}}, the anisotropy in the drag coefficients for broad-side on and end-on orientation, as given by Eq.~\ref{eq:settlingvel}, results in a lateral drift as the spheroid sediments. If $\alpha$ is the angle that the spheroid makes with the direction of gravity, $\hat{e}\cdot\hat{g}=\cos{\alpha}$, then the center of mass of the spheroid will drift at an angle $\delta < \alpha$ determined by the aspect ratio of the spheroid. Balancing the gravitational and the drag forces (Eq.~\ref{eq:settlingvel}) the angle at which spheroid drifts can be calculated as
\begin{align}
    \tan\left({\alpha - \delta}\right) = \frac{C_\parallel}{C_\perp}\tan{\alpha}.
    \label{eq:drift}
\end{align}

Fig.~\ref{fig:inclined} shows the trajectories of the center of mass for spheroids oriented at a range of different angles $\alpha$ with respect to the direction of gravity. The color bar indicates the angle $\alpha$ which varies from $0$ to $90$ degrees; the limiting values indicate the end-on and broad-side on orientation of the sedimenting spheroid. In these two cases, the spheroid sediments vertically, along $\hat{g}$. For any intermediate angle, the center of mass drifts laterally. In the figure, the symbols are the data obtained from the simulation, while the continuous lines are predictions of Eq.~\ref{eq:drift}. A reasonable match between the simulation data and the analytical prediction may be seen. The maximum lateral drift occurs for the spheroid oriented at $45^{\circ}$ to the direction of gravity. With further increase in $\alpha$, the extend of lateral drift decreases and disappears for the broad-side on orientation. The figure also shows the the black, circular markers ($\vcenter{\hbox{\Large{$\bullet$}}}$) that show the position of the spheroid at the end of the simulations. This is also also shown in the inset of Fig.~\ref{fig:inclined}. It can be seen that the total distance traveled by the spheroid decreases as $\alpha$ goes from $0^{\circ}$ to $90^{\circ}$, with the maximum drift observed for $\alpha = 45^{\circ}$. An important point to note from Fig.~\ref{fig:inclined} is the scale of abscissa and ordinate. The scale on ordinate is an order of magnitude larger than the scale of abscissa, indicating that the drift due to inclination of the spheroid is rather small. Despite the smallness of the drift, the numerical simulations clearly capture the drift with reasonable accuracy.

The drift gets even weaker with a decrease in the aspect ratio of the spheroid. Fig.~\ref{fig:inclined30} illustrates the dependence of the trajectory of the center of mass of the spheroid for various aspect ratios. The markers represent the data obtained from simulation and the continuous lines are analytical predictions based on Eq.~\ref{eq:drift}. It can be clearly seen that a sedimenting sphere does not drift, but as the aspect ratio increases the drifting angle increases. Again, the difference in the scale of abscissa and ordinate may be noted. The simulations accurately capture the drift for small aspect ratio particles despite the smallness of the drift.

\subsection{Spheroid suspended in a shear flow}
\label{sec:jeffery}

\begin{figure*}
    \centering
    \begin{subfigure}[b]{0.49\textwidth}
            \includegraphics[trim = 50 50 100 0, clip, width=\linewidth]{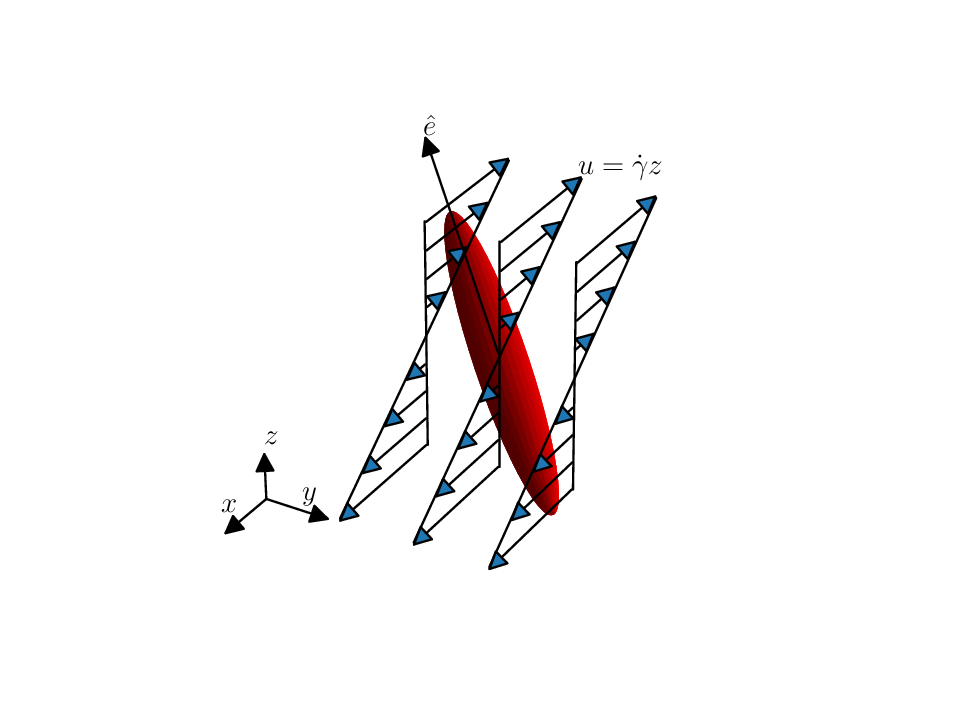}
            \caption{}
            \label{fig:spheroidinshear}
    \end{subfigure}
    \begin{subfigure}[b]{0.49\textwidth}
            \includegraphics[width=\linewidth]{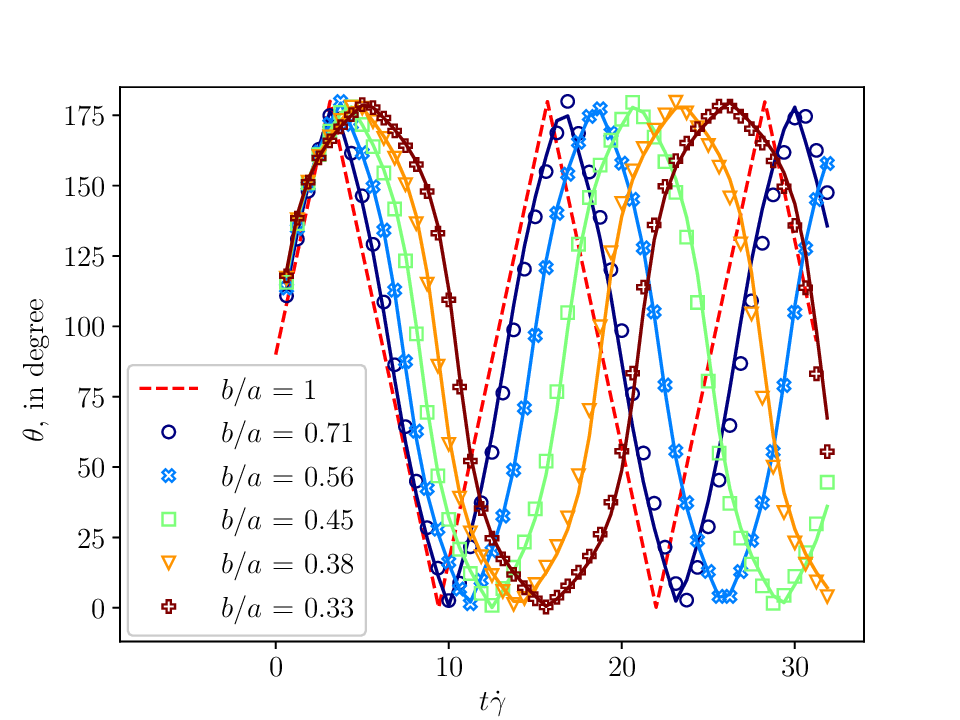}
            \caption{}
            \label{fig:jefferyaspect}
    \end{subfigure}        
    \begin{subfigure}[b]{0.49\textwidth}
            \includegraphics[trim = 110 30 50 40, clip, width=\linewidth]{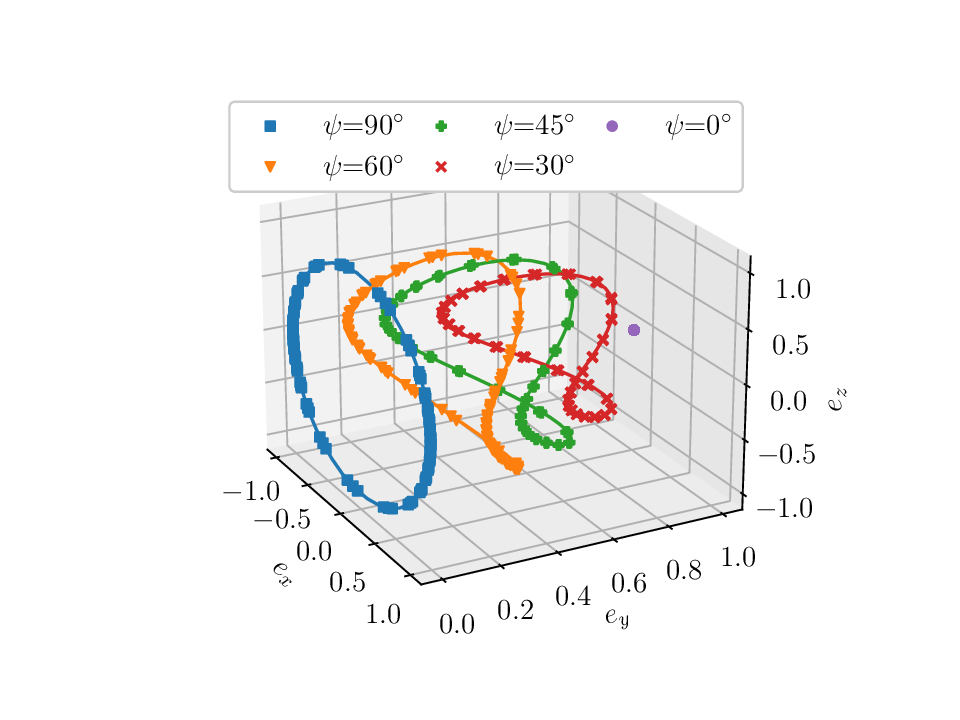}
            \caption{}
            \label{fig:jefferyoffplane}
    \end{subfigure}
    \begin{subfigure}[b]{0.49\textwidth}
            \includegraphics[trim = 0 0 0 0, clip, width=\linewidth]{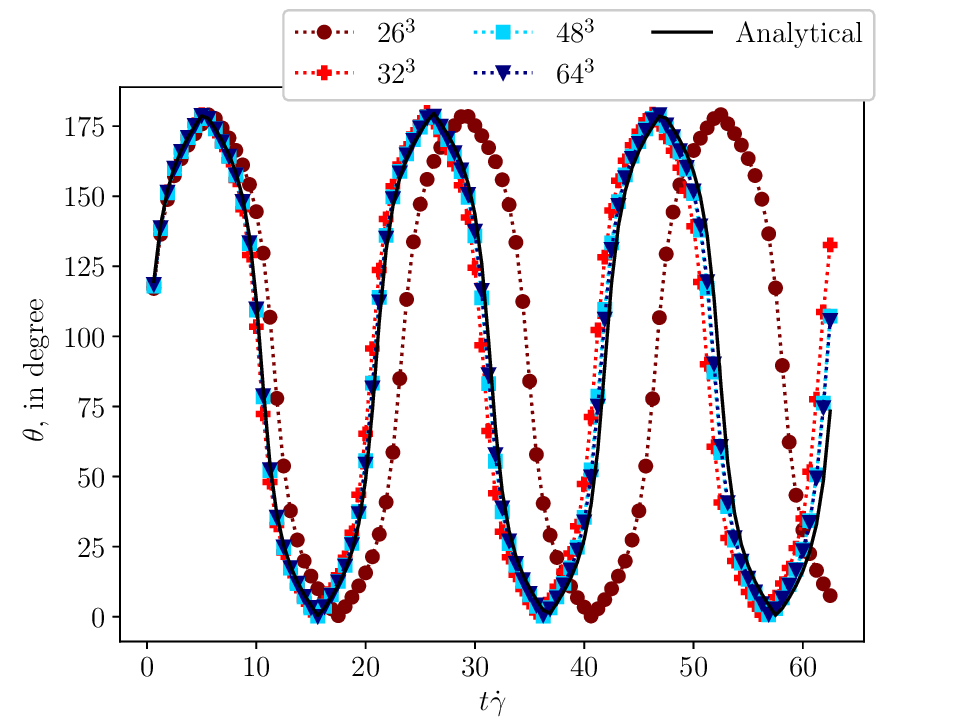}
            \caption{}
            \label{fig:jefferydomain}
    \end{subfigure}
    \caption{Jeffery orbits of spheroids in a simple shear flow at an imposed shear rate $\dot{\gamma} = 1.56 \times 10^{-5}$. (a) In-plane rotation of spheroids for various aspect ratios, $b/a$. The dashed line indicates the motion of an object (such as a sphere) rotating with the vorticity of the flow (angular velocity $= \frac{1}{2}$ of vorticity). (b) Out-of-plane rotation of the spheroid. Orientation of the spheroid described $e_x,e_y,e_z$ for the initial orientation $\phi=90,\theta=90,\psi$. The blue curve corresponds to the in-plane rotation. (c) In plane rotation (in flow-gradient plane) of the spheroid described by the Euler angle $\theta (t)$ in domains of different sizes.}
    \label{fig:jeffrey}
\end{figure*}

Due to the constraints imposed by low Reynolds number hydrodynamics, the sedimenting spheroid discussed in the previous section did not exhibit any rotational motion. Here, we consider a spheroid suspended in a simple shear flow. A sphere suspended in a simple shear flow rotates with an angular velocity commensurate with the vorticity of the imposed flow, but a spheroid exhibits even more complex, periodic motions as analytically calculated by \citet{jeffery1922motion}. As shown in Fig.~\ref{fig:spheroidinshear} consider a spheroid suspended in a simple shear flow $u = \dot{\gamma} z$ where $\dot{\gamma}$ is the imposed shear rate and $u$ is the velocity in the $y - $direction. As noted by \cite{jeffery1922motion, guazzelli2011physical} the time evolution of the orientation vector $\hat{e}$ is given by
\begin{align}
    \frac{d\hat{e}}{dt} = \boldsymbol{\Lambda} \cdot \hat{e} + \beta \left[ \bm{E}\cdot\hat{e} - \hat{e}(\hat{e} \cdot \bm{E} \cdot\hat{e})\right]
    \label{eqn:jeffery}
\end{align}
where $\bm{E}$ and $\boldsymbol{\Lambda}$ are the rate of strain and vorticity tensors, which are the symmetric and antisymmetric parts of the velocity gradient tensor respectively. $\beta = \frac{a^2 - b^2}{a^2 + b^2}$ is a measure of the aspect ratio of the spheroid. Note that $\beta = 0$ for a sphere.

Fig.~\ref{fig:jefferyaspect} shows the results obtained when the spheroid is placed in the flow-gradient plane. In this case, the symmetry of the configuration restricts the rotation of the spheroid to the flow-gradient plane and the orientation of the spheroid can be completely specified by the Euler angle $\theta$. In Fig.~\ref{fig:jefferyaspect}, temporal evolution of $\theta$ for spheroids of various aspect ratios are shown. The symbols are the data obtained from the simulations and the continuous lines are analytical predictions (Eq.~\ref{eqn:jeffery}). The dashed line indicates the rotation of a sphere ($b/a = 1$), an object that rotates with constant angular velocity. As the aspect ratio increases ($b/a$ decreases) (i) the angular velocity decreases as indicated by the longer time periods of revolution, and (ii) the angular velocity is not constant but varies as a function of time. Capturing both these features, a significant match between the simulation results and the analytical predictions in Fig.~\ref{fig:jefferyaspect} can be observed.

The suspended spheroid exhibits more complex, three dimensional trajectories when placed at an angle to the flow-gradient plane. The resulting trajectories, captured by the three components of the orientation vector, $e_x, e_y, e_z$, are shown in Fig.~\ref{fig:jefferyoffplane}. Different curves correspond to different initial orientations of the spheroid, specified by Euler angles $\phi = 90^{\circ}, \theta = 90^{\circ}, \psi$. Here, $\psi = 0$ shows the spheroid initially aligned along the vorticity axis. It then simply rotates with a constant angular velocity without any change in orientation, as indicated by a point in Fig.~\ref{fig:jefferyoffplane}. For any $0 < \psi < 90^{\circ}$, the spheroid exhibits three-dimensional trajectories. At $\psi = 90^{\circ}$ the spheroid will be in the flow-gradient plane and therefore the dynamics is restricted to a plane, as indicated by the circle. The symbols are obtained from the LB simulations and the continuous curve is the analytical prediction, Eq.~\ref{eqn:jeffery}. The simulations accurately capture these nontrivial, three-dimensional trajectories exhibited by the suspended spheroid in the simple shear flow.

The simple shear flow in lattice Boltzmann simulations is generated by placing two rigid walls in the $x - y$ plane, moving in opposite directions. However, Eq.~\ref{eqn:jeffery} is derived for a spheroid suspended in a fluid in an infinite domain. Therefore, to understand the role of hydrodynamic interaction of the spheroid with rigid walls and with periodic images, simulations were performed in domains of size $26^3$ up to $64^3$. The results are shown in Fig.~\ref{fig:jefferydomain}. The analytical prediction of Eq.~\ref{eqn:jeffery} is shown by the continuous line. It can be clearly seen that, compared to analytical predictions, the simulations show significant difference in smaller domains, but the results approach the analytical predictions with increase in domain size. The mismatch seen in smaller domains is solely due to the hydrodynamic interactions of the spheroid with its own images and the confinement imposed by the rigid walls.

\subsection{Spheroidal microswimmer}
\label{sec:squirmer}

\begin{figure*}
     \centering
     \begin{subfigure}[b]{0.3\textwidth}
         \centering
         \frame{\includegraphics[trim = 20 190 100 235, clip, width=\textwidth,angle=-90]{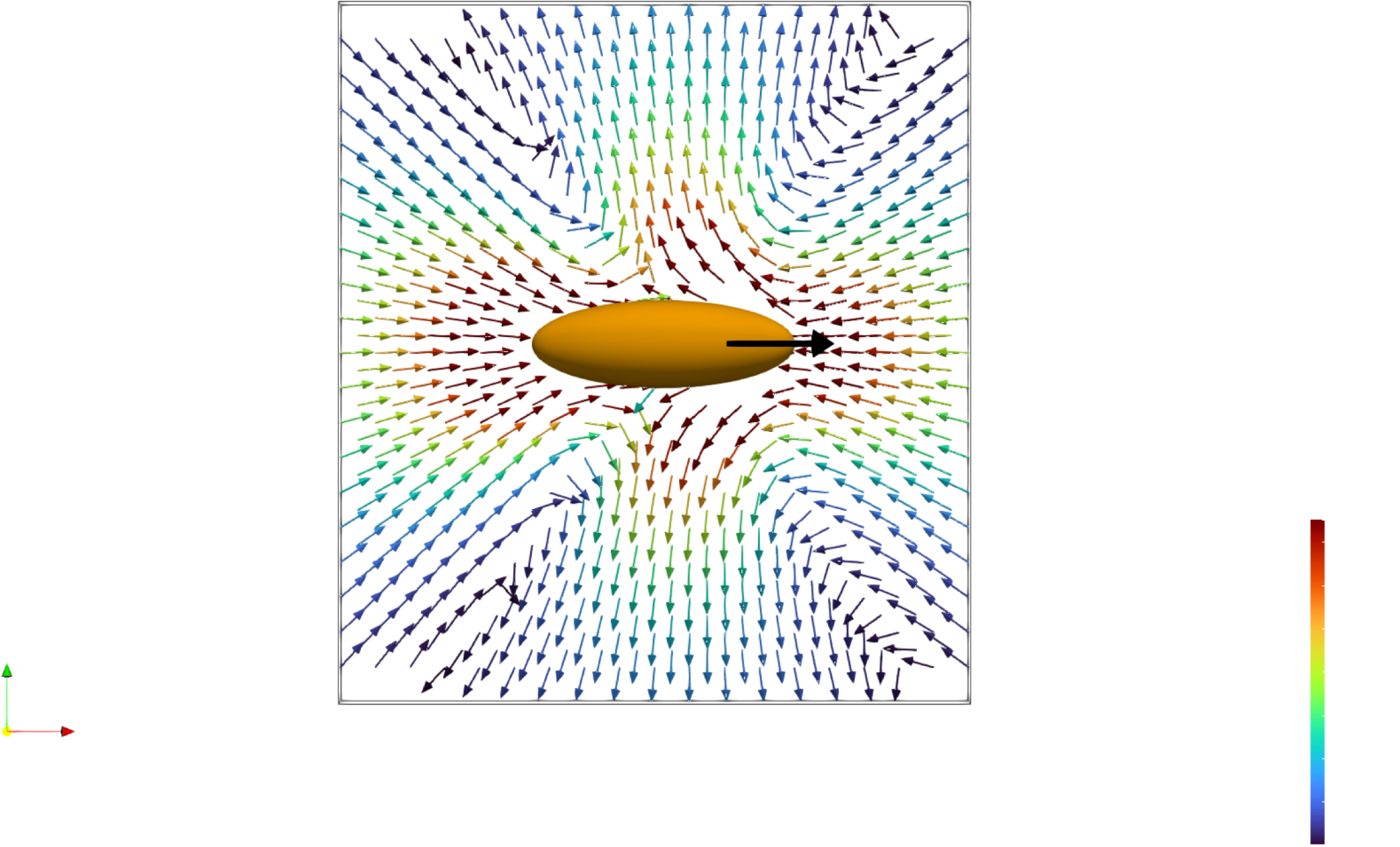}}
         \caption{}
         \label{fig:puller}
     \end{subfigure}
     \hfill
     \begin{subfigure}[b]{0.3\textwidth}
         \centering
          \frame{\includegraphics[trim = 20 190 100 235,  clip, width=\textwidth,angle=-90]{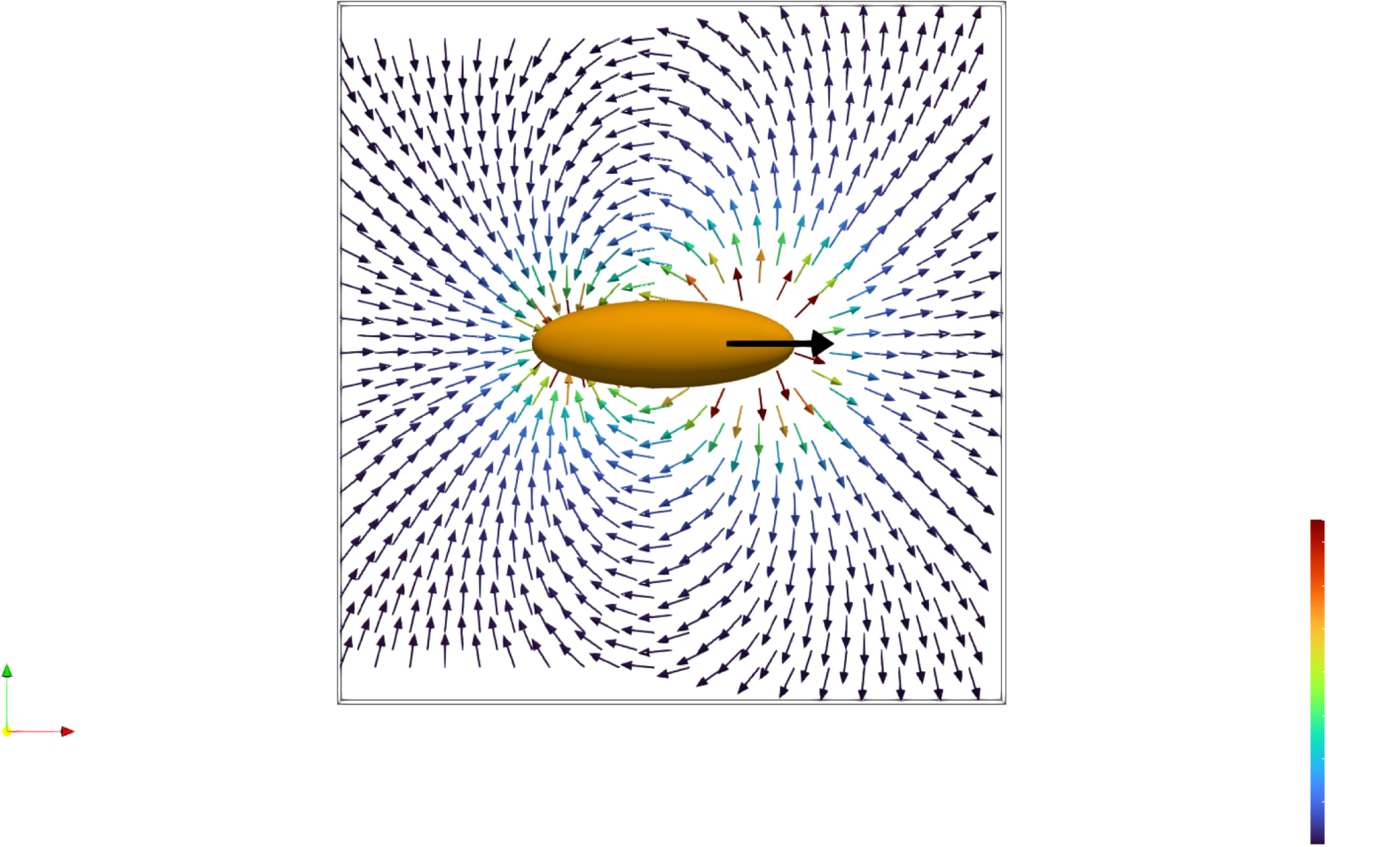}}
         \caption{}
         \label{fig:neutral}
     \end{subfigure}
    \hfill
     \begin{subfigure}[b]{0.3\textwidth}
         \centering
          \frame{\includegraphics[trim = 20 190 100 235,  clip, width=\textwidth,angle=-90]{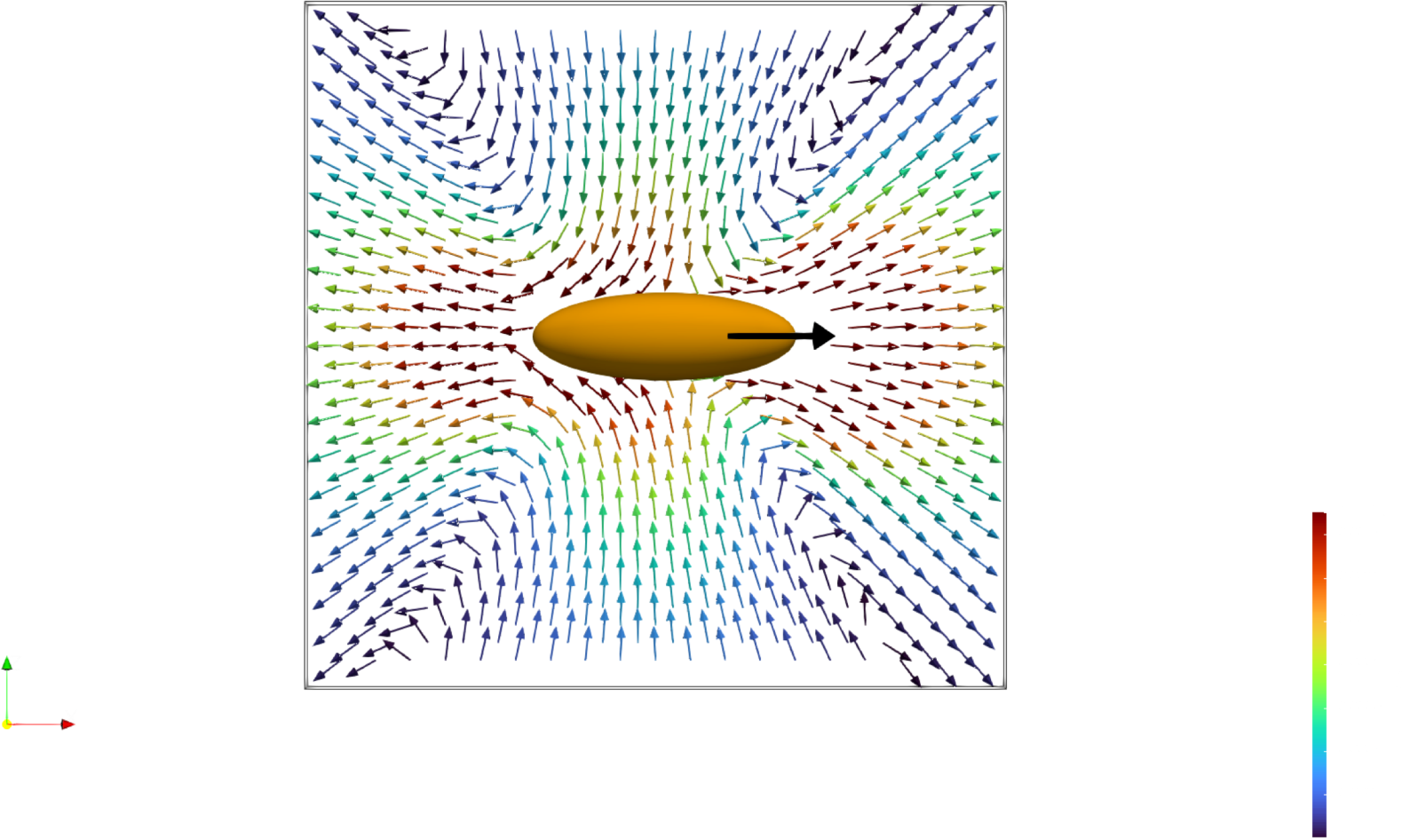}}
         \caption{}
         \label{fig:pusher}
     \end{subfigure}
        \caption{Fluid velocity developed around a spheroidal squirmer ($a > b = c$) with three different swimming strengths (a) $B_1 = B_2$, a puller (b) $B_1 \neq 0, B_2 = 0$, a neutral swimmer and (c) $B_2 = -B_1$, a pusher. The orientation of the squirmer is indicated with a thick, black arrow. The fluid velocity vectors are colored using a 'jet' color map such that red arrows indicate larger velocity than blue arrows.}
        \label{fig:velsquirm}
\end{figure*}

\begin{figure*}
    \centering
    \begin{subfigure}[b]{0.49\textwidth}
            \includegraphics[trim = 0 0 0 0, clip, width=\linewidth]{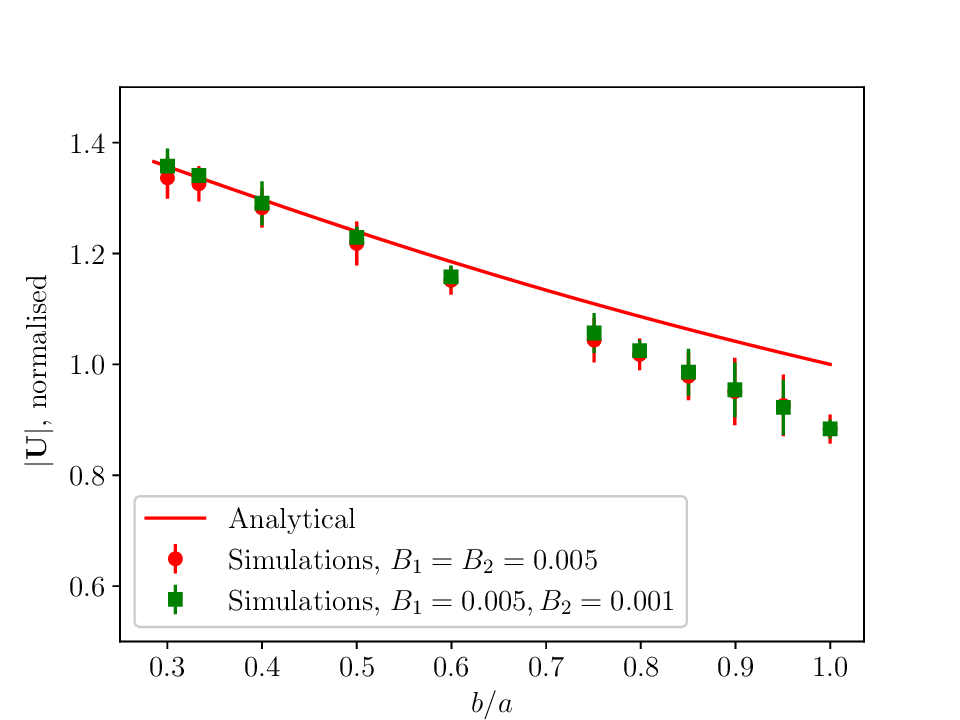}
            \caption{}
    \end{subfigure}     
    \begin{subfigure}[b]{0.49\textwidth}
            \includegraphics[trim = 0 0 0 0, clip, width=\linewidth]{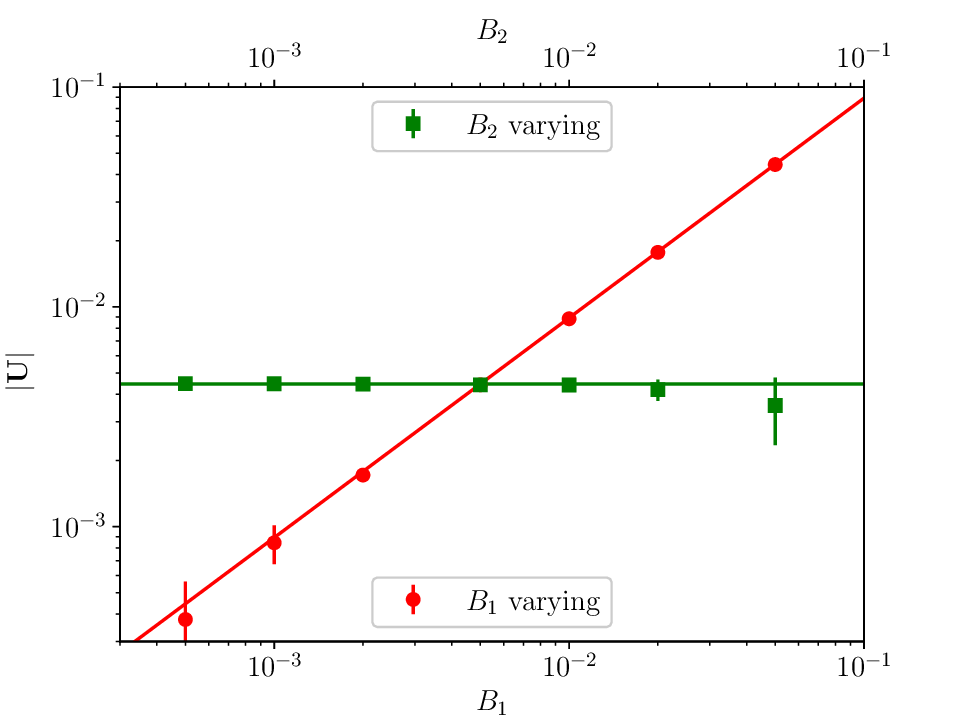}
            \caption{}
    \end{subfigure}     
    \caption{(a) Variation in the translational velocity of a spheroidal squirmer with respect to the changes in aspect ratio. The translational velocity $|\bm{U}|$ is normalized with that of a spherical squirmer $|\bm{U}| = \frac{2}{3}B_1$. (b) Variation in the translational velocity of a spheroidal squirmer as a function of $B_1$ (on the primary $x-$ axis) and $B_2$ (on the secondary $x-$ axis). Here, the continuous line is the analytical prediction and the symbols are data obtained from the simulations.}
    \label{fig:squirmer}
\end{figure*}

In the preceding sections we considered a rigid, passive spheroid responding to externally imposed forces or flow fields. In this section, we consider a spheroidal squirmer microswimmer, an active particle with slip boundary conditions that generates its own flow field and exhibits self-propulsion.

Following \citet{theers2016modeling} the spheroidal squirmer exhibits a surface slip velocity
\begin{align}
    \bm{u}_s = -B_1 (\bm{s} \cdot \hat{e}) \bm{s} - B_2 \zeta (\bm{s} \cdot \hat{e}) \bm{s} .
    \label{eq:squirmer}
\end{align}
$B_1$ and $B_2$ describe the strength of two modes of swimming: $B_1$ mode describes the swimmer as a source dipole and imparts the swimmer polarity while $B_2$ mode is apolar and describes the swimmer as a force dipole, the leading order description of a force-free particle. The latter distinguishes a pusher swimmer ($B_2 < 0$) from a puller swimmer ($B_2 > 0$). For $B_2 = 0$ the microswimmer is neutral.
In Eq.~\ref{eq:squirmer}
\begin{align}
\bm{s} = -\frac{\sqrt{a^2 - z^2}}{\sqrt{a^2-\epsilon^2z^2}}\hat{e} + \frac{\sqrt{1 - \epsilon^2}}{\sqrt{a^2-\epsilon^2z^2}}z\hat{e}_{\perp}     
\end{align}
is the surface tangent vector when the long axis of the spheroid $\hat{e}$ is oriented along the $z-$ axis of the coordinate system \cite{keller1977porous}. The spheroidal coordinate $\zeta = \left( \sqrt{x^2+y^2+(z+a\epsilon)^2} - \sqrt{x^2 + y^2 + (z-a\epsilon)^2} \right)/(2a\epsilon)$. In the absence of any other objects or confinement, the spheroidal squirmer translates with a swimming speed
\begin{align}
    U_s = B_1 \epsilon^{-1}\left(\epsilon^{-1} - (\epsilon^{-2} - 1)\coth^{-1}{(\epsilon^{-1})}\right).
    \label{eq:squirmervel}
\end{align}

The swimming velocity of a free, unconfined spheroidal squirmer depends only on the value of $B_1$ and not on the strength of the force dipole $B_2$. The same is true for a spherical squirmer.

In the lattice Boltzmann implementation, the slip velocity prescribed by Eq.~\ref{eq:squirmer} needs to be incorporated into the calculation of the velocity at the boundary nodes. A simple modification of  Eq.~\ref{eq:ub} allows
\begin{align}
\bm{u}_b = \bm{U} + \boldsymbol{\varOmega} \times (\bm{x}_b - \bm{x}_c) + \bm{u}_s (\bm{x_b}).
    \label{eq:ubsquirmer}
\end{align}
Apart from this aspect, the procedure outlined in section~\ref{sec:method} remains the same.

The fluid velocity fields generated by puller, neutral and pusher spheroidal microswimmers are shown in Fig.~\ref{fig:velsquirm}(a) - (c), respectively. The arrows represent the velocity vectors and the color code indicates the magnitude of the velocity ranging from blue as the lowest to red as highest velocity in the domain. The thick black arrow indicates the orientation of the swimmer. The neutral swimmer exhibits a flow field similar to that of a source-sink dipole aligned with the orientation of the spheroidal swimmer. The correspondence between the velocity fields of a puller and a pusher, namely fluid drawn from the front and back for the puller, or fluid pushed away from the front-back for the pusher, are also clearly evident in the figures.

Crucial for the validation of the implemented lattice Boltzmann algorithm is the swimming velocity of the spheroidal squirmer. The slip velocity results in the thrust force on the squirmer, while the viscous drag acts simultaneously in the opposite direction, resulting in a steady motion of the squirmer with the translational velocity given by Eq.~\ref{eq:squirmervel}. Hence, the steady swimming speed of the spheroidal squirmer at various aspect ratios $b/a$ is determined from the simulations for two cases, (i) when $B_1/B_2 = 5$, and (ii) when $B_1/B_2 = 1$. The results are shown in Fig.~\ref{fig:squirmer}(a). In this figure, the $y-$ axis is normalized by the steady translational velocity of a spherical squirmer, $\frac{2}{3}B_1$. The analytical solution given by Eq.~\ref{eq:squirmervel} is also plotted as a continuous line. The increase in the translational velocity of the squirmer with increasing aspect ratio (decrease in $b/a$) is clearly obtained in the simulations as predicted by the analytical calculations. The improvement in accuracy with decreasing $b/a$ is related to an effective increase in resolution as $b/a$ decreases ($a$ is varied, $b$ is fixed at 2.5 in the simulations).

For a further check of the dependency of these results on the choice of parameters $B_1$ and $B_2$, simulations were performed by varying these quantities over an order of magnitude. The results are shown in Fig.~\ref{fig:squirmer}(b). In this figure the primary and secondary $x-$ axes show $B_1$ and $B_2$, respectively, while the continuous lines show the analytical predictions. On varying $B_1$ the translational velocity of the squirmer linearly increases while variation in $B_2$ does not result in any changes in the translational velocity. Both observations agree quantitatively with the analytical predictions (Eq.~\ref{eq:squirmervel}). 

\begin{figure*}[htbp!]
     \centering
     \begin{subfigure}[b]{0.325\textwidth}
         \centering
         \includegraphics[width=\textwidth]{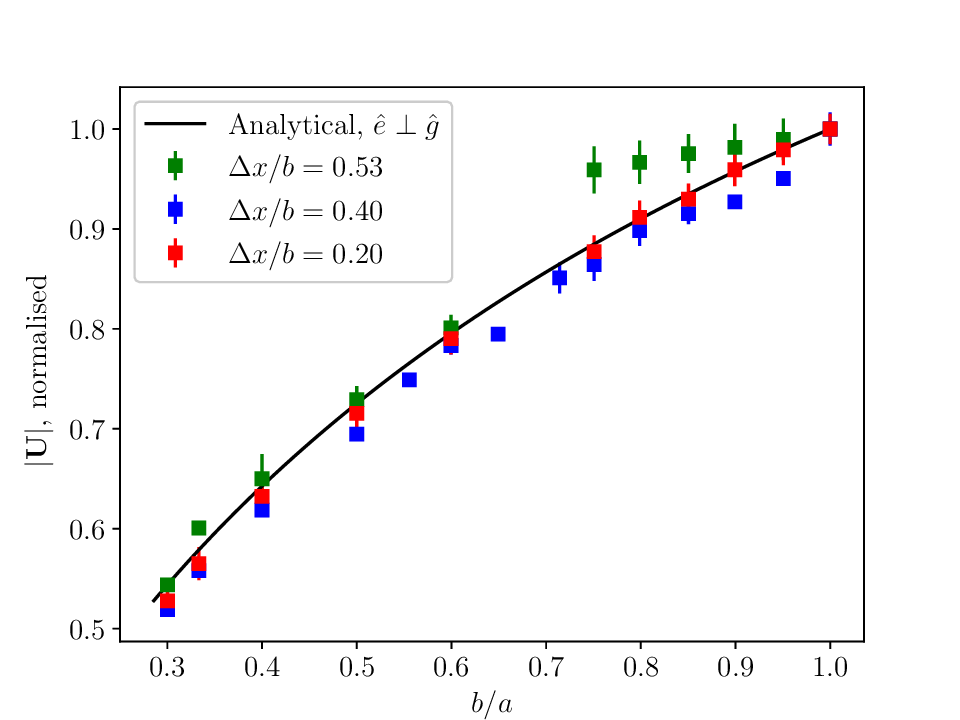}
         \caption{}
     \end{subfigure}
     \hfill
     \begin{subfigure}[b]{0.325\textwidth}
         \centering
         \includegraphics[width=\textwidth]{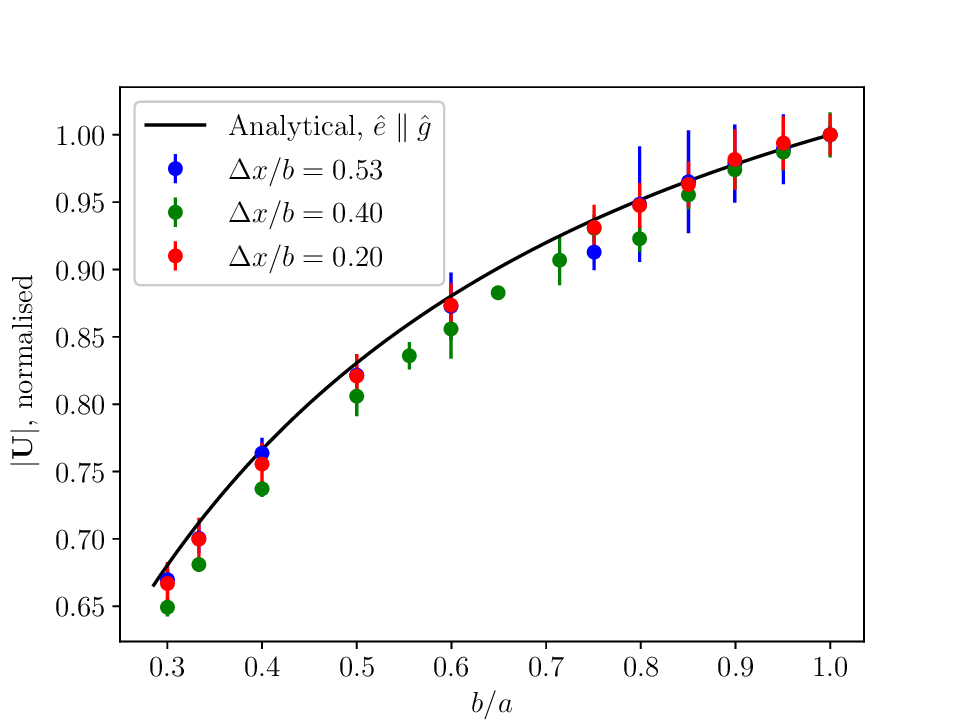}
         \caption{}
     \end{subfigure}
     \hfill
     \begin{subfigure}[b]{0.325\textwidth}
         \centering
         \includegraphics[width=\textwidth]{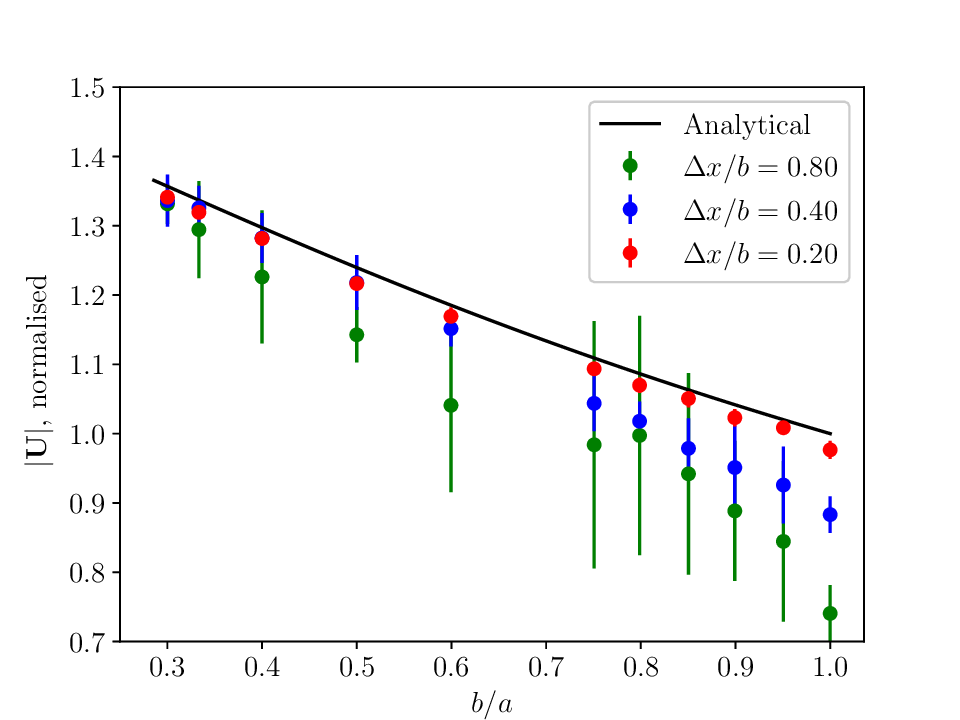}
         \caption{}
     \end{subfigure}
        \caption{\spt{The accuracy of the simulations is improved by changing the spatial resolution, $\Delta x / b$, illustrated for (i) a settling spheroid with the long axis of the spheroid oriented  perpendicular to the direction of gravity, i.e. $\hat{e} \perp \hat{g}$, (ii) a settling spheroid with the long axis of the spheroid oriented parallel to the direction of gravity, i.e. $\hat{e} \parallel \hat{g}$ and (iii) a self-propelling spheroidal squirmer. In (a) and (b) the kinematic viscosity is $\nu = 0.1$, whereas it is $\nu=0.167$ in (c). In all cases the translational velocity $|\bm{U}|$ is normalized with the translational velocity of a spherical particle of radius $r=b$. The error bars indicate the error in the calculations originating from the discrete shape of the particle, and are smaller than the symbol in most cases with higher resolution.}}
        \label{fig:resolution}
\end{figure*}

\spt{\subsection{Improving accuracy of results}
\label{sec:resolution}
In the previous subsections we discussed the results obtained from numerical simulations of spheroidal particles in different contexts. The generality of the method was also established by comparing the results from two different lattice Boltzmann schemes, namely the $D3Q19$ and $D3Q27$ model. In this section, and for completeness, we briefly discuss how an increased spatial resolution improves the accuracy of the results.\\
Fig.~\ref{fig:resolution} illustrates the results obtained from the numerical simulations of (i) a sedimenting spheroid in broad-side on configuration, (ii) a sedimenting spheroid in end-on configuration, and (iii) an active spheroidal microswimmer. In order to change the spatial resolution, the size of the particle was changed by keeping $\Delta x$ fixed, i.e., by changing $\Delta x/b$. Simultaneously, in (i) the domain size was also changed in order to avoid the changed effect of images (arising from periodic boundary conditions), and in (ii) the gravitational force for the settling particles and the slip velocity for the microswimmer are changed to maintain the same Reynolds number, the only non-dimensional number of relevance in the present investigations. In all cases the translational velocity of the spheroidal particle is determined and compared to the analytical solution. The plots clearly indicate that the numerical results approach the analytical solution as the spatial resolution is increased (i.e. smaller ratio $\Delta x/b$). However, this improved accuracy comes at an increased cost of computational resources and the parameters reported in the previous sections are recommended as a compromise between accuracy and computational resources.}

\section{Conclusions}
\label{sec:conclusions}

In this work we present a lattice Boltzmann algorithm to describe the modeling and hydrodynamic behavior of ellipsoidal particles. The lattice Boltzmann method is a reliable computational tool to investigate a variety of complex fluids. Moreover, it is highly scalable and suitable for simulating complex geometries. A simple bounce back scheme is implemented on the surface of the ellipsoid. The application of boundary conditions on the boundary nodes as done in this work makes the scheme easier to adapt for active particles which assume a slip-boundary condition. Similarly, regarding the ellipsoid as a solid body avoids defining other field parameters (say, order parameter field for microstructured fluids) on the solid nodes. The force and torque calculated on the boundary nodes are used to update the position and orientation of the ellipsoidal particles.

Determining the evolution of the orientation of the ellipsoidal particles is the most intricate part of the algorithm. To this end quaternions are used as they form the most efficient way and stable option for integrating the orientational degrees of freedom. As the definition of the quaternion is based on the angular velocity of the particle, it (i) prevents renormalization of errors and (ii) avoids separate numerical integration in different parts of the algorithm. Moreover, the use of quaternions permits determining the instantaneous moment of inertia and its time derivative without applying numerical approximations. Following \cite{nguyen2002lubrication} an implicit numerical scheme is also proposed to determine the instantaneous translational and angular velocity of the ellipsoidal particles in the fluid.

The method presented is validated using several known analytical solutions in low Reynolds number hydrodynamics. The translational velocity of a sedimenting spheroid in both broad-side on and end-on orientation compares very well with the analytical predictions. A sedimenting spheroid inclined at an angle to gravity maintains its orientation during the simulation time in accordance with the predictions of Stokes flow while its center of mass drifts at an angle intermediate between the direction of gravity and the orientation of the spheroid. The simulations captured this weak drift reliably. The applicability of the algorithm was found to be independent of the choice of simulation parameters, orientations and lattice Boltzmann models ($D3Q19$ and $D3Q27$).

Simulations were also performed to investigate the capability of the algorithm to capture Jeffery orbits accurately. These complex, three dimensional trajectories are traced by a spheroid when subject to simple shear flow. The results discussed in section~\ref{sec:jeffery} show a very good match with analytical predictions despite the significant complexity of the trajectories. Furthermore, the spheroidal particle exposed to a slip velocity to simulate the dynamics of microswimmers in a fluid. It was demonstrated that the algorithm can simulate different types of swimmers such as pushers, pullers and neutral swimmers. The swimming velocity of the squirmer and its variation with the aspect ratio of the squirmer and strength of various swimming modes match the analytical predictions. Currently we have restricted our analysis and validations to prolate spheroidal particles, but the implemented method is not restrictive; and future investigations will include oblate spheroidal and non-axisymmetric ellipsoidal particles.

Anisotropic particles are commonly observed in various areas of soft matter and complex fluids. Even in Newtonian fluids non-spherical particles exhibit rich dynamics as outlined in this work. Considering that the lattice Boltzmann method proves to be a reliable computational tool for simulating different types of complex fluids, including suspensions, emulsions, liquid crystals, this work shows that the LBM is also a very promising candidate for simulating hyper-complex liquids such as ellipsoidal particles dispersed in complex fluids.

 \section*{Acknowledgments}
Computational facilities at IIT Madras are duly acknowledged. SPT acknowledges the project support by I-Hub Foundation for Cobotics (IHFC), IITD and the travel support by IoE, IIT Madras.  
This work used the ARCHIE-WeSt High Performance Computer (\href{https://www.archie-west.ac.uk}{www.archie-west.ac.uk}) based at the University of Strathclyde.
For the purpose of complying with UKRI's open access policy, the authors have applied a Creative Commons Attribution (CC BY) license to any Author Accepted Manuscript version arising from this submission.
  
\bibliographystyle{apsrev4-1} 
\bibliography{biblio} 

\appendix
\subsection{\spt{Appendix: Algorithmic steps}}
\label{appA}
\spt{Below we provide the algorithmic steps related to the rigid body dynamics during one iteration step $t\to t+\Delta t$.
\begin{enumerate}
\item Perform lattice Boltzmann collision operation. \label{it:colln}
\item Perform lattice Boltzmann streaming operation on all fluid nodes except those streaming to boundary nodes.
\item Compute the velocity-independent force and torque $\mathbf{F}_0$ and $\mathbf{T}_0$ using Eq.~\ref{eq:F0}-\ref{eq:T0} based on post-collision distributions
\item Compute the moment of inertia tensor and its time derivative in the laboratory frame of reference using Eqs.~\ref{eq:Irot} and \ref{eq:Irotdiff}, respectively.
\item Solve the set of linear equations described by Eq.~\ref{eq:Fdiscrete}-~\ref{eq:Tdiscrete} to determine the translational ($\mathbf{U}(t+\Delta t)$)  and angular velocity ($\boldsymbol{\Omega}(t+\Delta t$) of the ellipsoid.
\item Perform mid-grid bounce back based on Eq.~\ref{eq:ub} for populations streaming from fluid to boundary nodes.
\item Update the position ($\mathbf{x}(t+\Delta t)$) using Eq.~\ref{eq:posupdate} and orientation in terms of quaternions ($\mathbf{q}(t+\Delta t)$) using Eq.~\ref{eq:qnplus0}-\ref{eq:qnplus1}.
\item Remap the nodes as solid, fluid and boundary nodes based on the updated position and orientation of the ellipsoid using Eq.~\ref{eq:ellipsoidsurface}. 
\item Re-compute the drag coefficient matrices using Eq.~\ref{eq:drag1}-\ref{eq:drag4}.
\item Goto step \ref{it:colln}.
\end{enumerate}
}

\end{document}